\documentclass[review]{elsarticle}

\usepackage{graphicx}
\usepackage{array}
\usepackage{amssymb}
\newcolumntype{M}[1]{>{\centering\arraybackslash}m{#1}}
\usepackage{caption}
\usepackage{subcaption}
\usepackage{amsmath}
\usepackage{multirow}
\usepackage{multicol}
\usepackage[version=4]{mhchem}
\usepackage{lineno,hyperref}
\modulolinenumbers[3]
\usepackage{setspace}
\usepackage{longtable}
\usepackage{xcolor}
\journal{TBD}

\begin{document}

\begin{frontmatter}

\title{Shape Characterization of Ferrous Burden Material of Blast Furnace Feed using Image Analysis}

\author[tsl]{Arijit Chakrabarty}
\author[iitk]{Aman Tripathi}
\author[iitk]{Vimod Kumar}
\author[iitk]{Anurag Tripathi\corref{corr}}
\author[iitk]{Arpit Agarwal}
\author[iitk]{Vishwaraj Singh}
\author[tsl]{Saprativ Basu}
\author[tsl]{Samik Nag}
\cortext[corr]{Corresponding Author\\
E-mail Address: anuragt@iitk.ac.in\\
Postal Address: Department of Chemical Engineering, Indian Institute of Technology Kanpur, Kanpur, India - 208016.}

\address[tsl]{R$\&$D Division, TATA Steel, Jamshedpur, India - 831001.}
\address[iitk]{Department of Chemical Engineering, Indian Institute of Technology Kanpur, Kanpur, India - 208016.}

\begin{abstract}

In this study, we attempt to characterise the shape of three different types of grains commonly used in the iron and steel-making industry, namely pellet, sinter and iron ore lump. We choose particles over the entire size ranges used in industrial-scale blast furnace and consider two different size ranges of pellet particles and four different size ranges for sinter and iron ore lumps. 
We perform image analysis to calculate size and shape-related properties of the grains. We select some of the common length scales used to measure the size of the particles and categorise them in different classes. 
We show that the length scales of a particular category class are well correlated with each other. We identify the independent, uncorrelated length scales for the particles from different categories. 
Using these uncorrelated length scales, we define different shape descriptors and obtain the distribution of these shape descriptors for each component of the blast furnace feed. 
Our image analysis results show that the cumulative distribution curves for these shape descriptors turn out to be nearly independent of the size range for a given type of material. Our study identifies the three key shape descriptors that are required to characterize the shape of the blast furnace feed.
Two of these shape descriptors, namely the aspect ratio and the circularity, have been considered important by the researchers earlier as well. The third shape descriptor, the average contact eccentricity to the projected particle diameter ratio, usually not considered to be an important shape descriptor in previous studies, is of high relevance for Discrete Element Method simulations of granular materials. 
\end{abstract}

\end{frontmatter}


\section{Introduction}

Handling and processing of granular materials has widespread applications in various industries like pharmaceuticals, food processing, mining and metals, etc. Granular mixtures are composed of grains that differ in size, shape, density, etc. and hence their flow and transportation often result in unwanted segregation. Such separation of granular materials into regions of different properties can pose serious challenges in applications which require a well-mixed product. 
While the segregation due to differences in size and/or density has been experimentally as well as theoretically studied a lot, the shape of the particles in such theoretical studies has always been assumed to be spherical. For non-spherical grains, the characterisation of both size as well as shape becomes a challenge.  
A relevant example of these types of non-spherical grains of special industrial importance is from the iron and steel industry. The steel makers around the world commonly employ the blast furnace route of iron-making which utilises iron ore as the raw material and smelts it to produce liquid iron which is sent to steel-making shops for finishing. Quite a significant portion of the iron ore obtained from mines is in the form of fines (particle size smaller than 1 mm) while the rest  is in the form of iron ore lumps. Direct usage of such small fines in the blast furnace leads to choking and upheavals in the operational stability. To avoid this, the fines are first processed and converted into larger size agglomerates of two different types, namely, sinter and pellets.
The ferrous burden materials charged into the blast furnace comprises of these pellets and sinter as well as the iron ore lumps (see fig. 1).
 
While the major constituent of these materials is iron ore, there are significant differences in their chemistry as well as their physical nature. Pellets are largely spherical in shape, whereas sinter shape is more uneven with a very irregular surface topography and has high porosity compared to the other two components. Iron ore lumps partially resemble polyhedral particles and may have sharp edges. Due to their spherical shape, pellets tend to flow more easily and have a lower angle of repose compared to the sinter and iron ore lumps. 
Size distribution of sinter and iron ore lumps is significantly wider than pellets, making the ferrous feed more susceptible to size segregation. Changes in the size and shape of the ferrous burden material charged in the furnace has significant influence on the porosity and packing of the resulting material bed, leading to changes in the furnace performance. 
Tracking and identifying sources of such changes, however, requires the characterisation of granular feed using a consistent and reliable method.

Quantitative description of the size and shape of the constituent particles is crucial to understand their effect on distribution of different types of materials and their subsequent packing inside the blast furnace. These bed properties ultimately determine the gas flow dynamics through the bed that largely dictates the operational stability and efficiency of the blast furnace.

\begin{figure}[htbp]
\begin{center}
\includegraphics[width=0.3\textwidth]{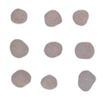}
\includegraphics[width=0.31\textwidth]{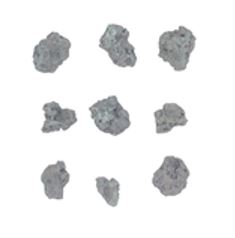}
\includegraphics[width=0.32\textwidth]{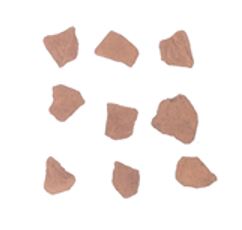}
\caption{(L-R) Pellets, sinter and iron ore lumps.}
\label{particles}       
\end{center}
\end{figure}

The characterisation of the size distribution of granular feed material is typically done using
sieve analysis. The characterisation of the shape of the grains, however, remains a difficult
task. Shape characterisation of such particles has been historically done using dimensionless parameters called shape factors or shape descriptors. These shape descriptors are usually dimensionless combinations of  length-scales and/or areas and/or volumes related to the particle. 
Ideally, a $3$D analysis of the particles is more comprehensive as all topographical features of a particle can be captured in this case. However, $3$D contour capturing techniques are relatively more time consuming. 
Due to this reason, $2$D imaging for size and shape analysis is preferred by most researchers. These $2$D shape parameters are determined from detailed image analysis of particles whose size can range from few microns to few millimetres.
Application of 2D imaging has been successfully employed for materials from a vast variety of fields such as mineral and metal powders \cite{hentschelpage, mikli, garboczi, lecoq, kulu}, particles subjected to wear and wear debris \cite{pintaude, howe, nowell, kaya, frances, mohanty}, sand \cite{duris, lietal, xie, linzhuli, he}, volcanic media \cite{liu, andronico, coltelli, manga, nemeth1, dellino, jordan}, cement and concrete \cite{sinkhonde, bu, yue, mora}, bio-materials \cite{sinkhonde, bouwmann, zhengphd, souza, becke}, etc. 
For the 2D image analysis, high resolution images of the samples are taken. Various microscopy techniques \cite{mikli, frances, liu, andronico, zhengphd, grulke, belaroui1, vivier, pons3, maroof}, digital cameras \cite{mohanty, xie, mora, bouwmann, souza, leibrandt, roostaei, kroner, kuo, podczeck} as well as scanners \cite{duris, yue, bagheri, pons4} have been used for capturing the images. The captured images are typically converted into black and white images using algorithms that first change the colours of individual pixels into grayscale counterparts and then they are converted to either black or white depending on the threshold value decided. 
Particle projections are converted to either black or white on a background of the other colour. Using the projection pixel data, different parameters of the shaded region i.e., of the silhouette, can be calculated. Image conversion and shape descriptor calculations are done using customised codes or commercial packages like ImageJ \cite{imagej}, MATLAB Image Processing Toolbox \cite{matlab}, Sigma Scan Pro \cite{sigmascan}, Image-Pro \cite{improplus}, etc. While it is important that the imaging procedure be as intricate as possible to capture high resolution images, software may also play an important role in determining different shape descriptors. Duris et. al. \cite{duris} performed image analysis of scanner-taken images of sand particles using three different software, namely ImageJ, MATLAB and Sigma Scan Pro. They found that perimeter values of particle contours were always higher using Sigma Scan Pro than both ImageJ and MATLAB, both of which returned comparable values. 

From a $2$D contour, many parameters related to length and area can be defined and calculated. However, it remains a task to identify which shape descriptors are unique and capable of describing the particle shape. Hentschel and Page \cite{hentschelpage} analysed $11$ different shape descriptors for commercially manufactured powders and deduced that aspect ratio (for elongation) and root form factor (for ruggedness) are enough to describe the particle shape. Bouwmann et. al. \cite{bouwmann} deduced that circularity and projection shape factor work well for describing shape of granules of different bio-powders. Berrezueta et. al. \cite{berrezueta} formulated a new shape descriptor which could quantify particle shape, with it being a function of $4$ independent shape descriptors. Mora et. al. 
\cite{mora} deduced that only $2$ shape descriptors out of $6$ are enough to describe the roundness and angularity of concrete aggregates. Liu et. al. 
\cite{liu} conducted cluster analysis to deduce that $3$ shape parameters can provide a robust description of volcanic ash particle shape. Sinkhonde et. al. \cite{sinkhonde} were unable to correlate values of traditionally measured circularity and roundness for cement and different bio-material particles. Advanced studies \cite{xie, linzhuli, bagheri, han} have shown that shape descriptors formed using $3$D measurements can be correlated with those formed using $2$D measurements, thus bypassing the need of more involved $3$D measurements to some extent.

As mentioned before, shape descriptors are non-dimensional quantities represented as the ratio of relevant physical quantities of the particles relating to their size, area, volume etc. The choice of the physical properties used to characterize these shape descriptor also varies among different studies. 
For example, in order to characterize the length scales associated with a particle, at least three length scales (one corresponding to each direction) are needed for any particle. 
In the case of image analysis using 2D projection of the particles, minimum two length scales are needed. Many choices for these length scales exist and have been used by researchers. For example, many studies \cite{hentschelpage, garboczi, lecoq, frances, mohanty, xie, linzhuli, he, liu, andronico, coltelli, souza, becke, grulke, belaroui1, pons3, roostaei, singh, zeng, kulkarni, olson} use the minimum and maximum Ferret diameter as the two relevant length scales and use them to obtain one of the key shape descriptor, often referred to as the aspect ratio. 
The popularity of the Ferret diameter as the relevant length scale can be attributed to the measurement techniques used decades ago. Given the limited use of computers and image analysis, Feret diameter was one of the most practical ways of measuring particle size using microscopes or vernier callipers. With the availability of advanced image processing tools with modern computers, more detailed measures of the particle size are possible. For example, image processing tools   provide the major and minor axis of the ellipse having equal amount moment of inertia as the two length scales which are used by some researchers \cite{mikli, liu, singh, endoh, stienkijumpai} to describe the axial ratio (sometimes also called anisometry). 
A detailed account of various shape descriptors used in literature is listed in tables~\ref{sd_length_list},~\ref{sd_length_perimeter_list} and ~\ref{sd_length_perimeter_area_list}. 

Table~\ref{sd_length_list} refers to shape descriptors that have been largely formulated as ratios of one length scale to the other. For any $2$D shape, a variety of $1$D parameters can be defined to quantify particle size. A popular use of Ferret diameters is evident from Table ~\ref{sd_length_list}, where $F_{min}$ and $F_{max}$ are the minimum and maximum Ferret diameters of a particular shape. Sometimes, the mean Ferret diameter $F_{mean}$ is also considered. In addition to these, the dimensions of the minimum bounding box ($F_{mbb}$ and $F_{mbb90}$) of a $2$D shape have also been used. Largest inscribing circle diameter $D_i$ and smallest circumscribing circle diameter $D_c$ have found use as well. Recently, use of $a$ and $b$ which are the semi major and semi minor axes lengths of an ellipse having the same second moment of inertia as the oriented $2$D shape has also gained prominence. Some studies also calculate $R_{av}$ which is the average of multiple $R_i$s where $R_i$ is the length of the line connecting the centroid of the $2$D shape to the $i^{th}$ point along its contour.

\begin{center}
\begin{longtable}{m{3cm}m{3cm}m{5cm}}
	\hline
	\textbf{Shape Descriptor} & \textbf{References} & \textbf{Nomenclature Used}\\
        \hline
        $\frac{b}{a}$ & \cite{liu, endoh, stienkijumpai} & Axial Ratio \cite{liu}, Anisometry \cite{endoh}, Reciprocal Axial Ratio \cite{stienkijumpai}\\
        \hline
	$\frac{a}{b}$ & \cite{mikli, singh} & Aspect \cite{mikli}, Anisometry \cite{singh}\\
	\hline
        $\log_2 (\frac{a}{b})$ & \cite{mikli, kulu} & Elongation\cite{mikli, kulu}\\
	\hline
	$\frac{F_{min}}{F_{max}}$ & \cite{hentschelpage, garboczi, lecoq, frances, mohanty, xie, linzhuli, he, liu, andronico, coltelli, souza, becke, grulke, belaroui1, pons3, roostaei, pons4, singh, zeng, kulkarni, olson, pons2, pons1, belaroui2} & Aspect Ratio \cite{hentschelpage, mohanty, linzhuli, liu, andronico, coltelli, souza, becke, grulke, pons3, roostaei, zeng, kulkarni, olson, pons2}, Elongation \cite{lecoq, frances, xie, belaroui1}, Flatness \cite{he}, Elongation Factor \cite{singh}\\
        \hline
        	$\sqrt{\frac{D_i}{D_c}}$ & \cite{mikli, maroof, kroner, bagheri,zhou, endoh, wojnar} & Square Root of Irregularity Parameter \cite{mikli, zhou}, Inscribed Circle Sphericity \cite{maroof, kroner, bagheri}, Irregularity Parameter \cite{endoh}\\
        \hline
	$1 - \sqrt{\frac{1}{N - 1} \sum_{i=1}^{N} (\frac{R_i}{R_{av}} - 1)^2}$ & \cite{bouwmann} & Radial Shape Factor \cite{bouwmann}\\
        \hline
	$\frac{F_{mean}}{F_{max}}$ & \cite{hentschelpage, michalski} & \\
	\hline
        $\frac{F_{mbb}}{F_{mbb90}}$ & \cite{hentschelpage, yue, zhengphd} & \\
	\hline
\caption{List of shape descriptors involving different length scales used in literature.}
\label{sd_length_list}
\end{longtable}
\end{center}

\begin{center}
\begin{longtable}{m{3cm}m{3cm}m{5cm}}
	\hline
	\textbf{Shape Descriptor} & \textbf{References} & \textbf{Nomenclature Used}\\
        \hline
	$\frac{P}{\pi D_p}$ & \cite{mohanty, nemeth1, dellino, jordan, maroof, singh, zeng, kulkarni, endoh, durig, nemeth2} & Convexity \cite{mohanty}, Circularity \cite{nemeth1, dellino, jordan, singh, zeng, endoh, kulkarni, durig, nemeth2}, Perimeter Sphericity \cite{maroof}\\
	\hline
	$\frac{P_{ch}}{P}$ & \cite{liu, vivier, pons3, maroof, leibrandt, kroner, kulkarni, olson, endoh, liu2, michalski} & Convexity \cite{pons3, leibrandt, kulkarni, olson, endoh, liu2}, Roughness \cite{maroof}, Convexity Ratio \cite{kroner}\\
	\hline
	$\frac{\pi F_{mean}}{P}$ & \cite{hentschelpage, maroof, michalski} & Convexity \cite{hentschelpage}, Roughness Inverse \cite{maroof}\\
       \hline
        $\frac{2 \pi R_{av}}{P f} - \sqrt{1 - (\frac{b}{a})^2}$ & \cite{bouwmann, podczeck, zeng} & \\
	\hline
	$\frac{P}{D_{p}}$ & \cite{mikli} & Roundness Factor \cite{mikli}\\
	\hline
	$\frac{P}{\pi F_{max}}$ & \cite{hentschelpage, liu} & Convexity Feret \cite{liu}\\
	\hline
	$\frac{\pi F_{min}}{P}$ & \cite{hentschelpage} & \\
        \hline
\caption{List of shape descriptors used in literature that involve different length scales and perimeter.}
\label{sd_length_perimeter_list}
\end{longtable}
\end{center}

\begin{center}
\begin{longtable}{m{1.6cm}m{3cm}m{5cm}}
	\hline
	\textbf{Shape factor} & \textbf{References} & \textbf{Nomenclature Used}\\       
        \hline
	$\frac{2 \sqrt{\pi A}}{P}$ & \cite{hentschelpage, xie, linzhuli, bu, roostaei, olson} & Root Form Factor \cite{hentschelpage}, Sphericity \cite{xie, linzhuli, bu, roostaei}, Circularity \cite{olson},\\
        \hline
	$\frac{P}{2 \sqrt{\pi A}}$ & \cite{liu, sinkhonde, berrezueta} & Circularity \cite{liu, sinkhonde}, Normalized Perimeter vs. Area Ratio \cite{berrezueta}\\
        \hline
	$\frac{2 \sqrt{A}}{\sqrt{\pi} F_{max}}$ & \cite{hentschelpage, souza, grulke, pons3, pons1} & Compacity \cite{souza}, Compactness \cite{grulke, pons3}\\
    \hline
	$\frac{A}{A_{ch}}$ & \cite{lecoq, mohanty, linzhuli, liu, sinkhonde, mora, belaroui1, vivier, pons3, leibrandt, roostaei, kroner, kuo, pons4, zhou, berrezueta, kulkarni, olson, endoh, pons2, liu2} & Concavity Index \cite{lecoq, vivier, pons4, pons2}, Solidity \cite{mohanty, liu, sinkhonde, pons3, leibrandt, kroner, kulkarni, olson, endoh, liu2}, Convexity \cite{linzhuli, roostaei, zhou}, Convexity Ratio \cite{mora}, Concavity Factor \cite{belaroui1}, Fullness Ratio squared \cite{kuo}, Solidity-Convexity \cite{berrezueta}\\
    \hline    
    $\frac{4A}{\pi {F_{max}}^{2}}$ & \cite{mohanty, duris, lietal, liu, sinkhonde, pons3, kroner, podczeck, kulkarni} & Roundness \cite{mohanty, duris, liu, sinkhonde, pons3, kroner, kulkarni}, Circularity \cite{lietal}, Projection Sphericity \cite{podczeck}\\
	\hline
	$\frac{4 \pi A}{P^{2}}$ & \cite{hentschelpage, kulu, pintaude, mohanty, duris, lietal, liu, andronico, coltelli, manga, yue, zhengphd, pons3, souza, maroof, leibrandt, kroner, podczeck, bagheri, berrezueta, singh, zeng, kulkarni, liu3} & Form factor \cite{hentschelpage, kulu, pintaude, mohanty, liu, andronico, coltelli, pons3, maroof, liu3}, Circularity \cite{duris, lietal, bouwmann, leibrandt, kroner, podczeck, bagheri, kulkarni}, Shape Factor \cite{manga, yue, zhengphd, berrezueta, zeng}, Roundness \cite{souza}\\
	\hline
	$\frac{P^{2}}{4 \pi A}$ & \cite{mikli, lecoq, howe, nowell, frances, he, belaroui1, pons3, kuo, pons4, singh, endoh, pons2, pons1, belaroui2, wojnar, montero} & Roundness \cite{mikli, he, kuo}, Circularity \cite{lecoq, frances, belaroui1, pons4, pons2, pons1, belaroui2, montero}, Roundness Factor \cite{howe, nowell}, Form Factor \cite{pons3}, Surface Factor \cite{singh}, Surface Roughness \cite{endoh}, Irregularity \cite{wojnar}\\
    	\hline
	$\frac{2 \sqrt{A}}{\sqrt{\pi} F_{mean}}$ & \cite{hentschelpage} & \\
	\hline
	$\frac{\sqrt{\pi} F_{min}}{2 \sqrt{A}}$ & \cite{hentschelpage} & \\
        \hline
	$\log_2 ({\pi a b})$ & \cite{mikli, kulu} & Dispersion \cite{mikli, kulu}\\
	\hline
\caption{List of shape descriptors used in literature that involve different length scales, perimeter and area.}
\label{sd_length_perimeter_area_list}
\end{longtable}
\end{center}

Table~\ref{sd_length_perimeter_list} lists shape descriptors which involve the perimeter of the silhouette along with a length scale. These shape descriptors also use the equivalent area diameter $D_p$ as an important length scale along with convex hull perimeter $P_{ch}$ in various studies. 
Table~\ref{sd_length_perimeter_area_list} lists shape descriptors which take into account different area scales, particularly projected area $A$ and convex hull area $A_{ch}$. 
Researchers have used different measures of particle length scales along with the perimeter and area to formulate shape descriptors of their choice.
Note that since the equivalent area diameter $D_p$ is calculated from the projected area $A$ of the particle, the shape descriptors defined using $D_p$ in table~\ref{sd_length_perimeter_list} are related to some of the shape descriptors defined using projected area $A$ in table~\ref{sd_length_perimeter_area_list}. Similarly some of the shape descriptors in the same table are also related to each other. We return to this point in a later section.

A quick look at the last column of these tables makes it evident that a particular shape descriptor has been given different names in different works. For example, the shape descriptor $F_{min}$/$F_{max}$ has been popularly described as aspect ratio by many researchers \cite{hentschelpage, garboczi, mohanty, linzhuli, liu, andronico, coltelli, souza, becke, grulke, roostaei, zeng, kulkarni, olson, pons2}, but some researchers have used other terms like elongation \cite{lecoq, frances, xie, belaroui1}, elongation factor \cite{singh} and flatness \cite{he} for the same. Similarly, convexity as a shape descriptor has been named as the ratio of convex hull perimeter $P_{ch}$ to perimeter $P$ \cite{pons3, leibrandt, kulkarni, olson, endoh, liu2} as well as for the ratio of area $A$ to convex hull area $A_{ch}$ \cite{linzhuli, roostaei, zhou}, but it has also been used to denote $P$/$\pi D_{p}$ \cite{mohanty} and $\pi F_{mean}$/$P$ \cite{hentschelpage}. 
Note that circularity has been used to denote shape descriptors like $P$/$\pi D_{p}$ \cite{nemeth1, dellino, jordan, singh, zeng, kulkarni, endoh, durig}, which is identical to $P$/$2\sqrt{\pi A}$ used by some researchers \cite{liu, sinkhonde}. However, the expression $4 \pi A$/$P^2$ has also been denoted as circularity by \cite{duris, lietal, bouwmann, leibrandt, kroner, podczeck, bagheri, kulkarni}, which is simply the squared inverse of that used by others mentioned before. 

We follow a systematic approach and suggest appropriate choices of the relevant shape descriptors in this work. In this study, we limit ourselves to characterise the shape of three different types of grains commonly used in iron and steel making industry, namely pellet, sinter and iron ore lump.
We attempt to classify the various length scales computed from the image analysis into different groups. We find that many of the silhouette properties of a particular group are very well correlated with other properties of the same group, regardless of the material type being considered. The organisation of the rest of the paper is as follows. Section 2 briefly describes the imaging methodology and various particle level properties measured using the image analysis. Section 3 reports the correlation among these particle properties and identifies the independent and uncorrelated particle level properties that can be used to define shape descriptors. Section 4 presents the distribution of various possible shape descriptors and their intercorrelation with each other for the ferrous burden material. Finally, the conclusion of this study are presented in section 5.


\section{Materials and Methods}

\begin{figure}[h!]
\begin{center}
\includegraphics[width=0.9\textwidth]{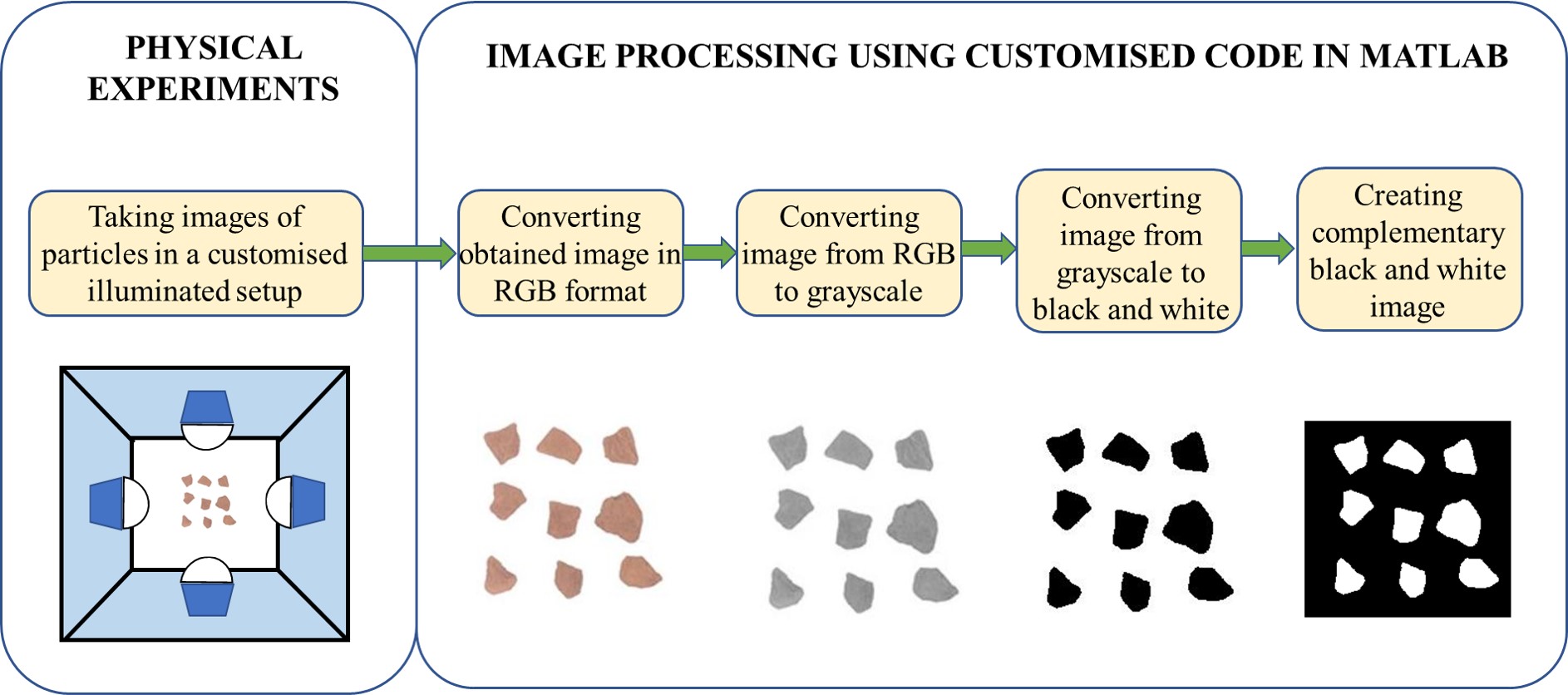}
\caption{Schematic describing the methodology used for 2D image analysis.}
\label{2d_imaging}       
\end{center}
\end{figure}

As mentioned before, we characterise the shape of pellet, sinter and iron ore lump commonly used in iron and steel making industry. 
While the pellet particles are nearly spherical in shape, the iron ore lumps and highly irregular sinter particles deviate significantly from the spherical shape. We have used two different size ranges of pellet particles (8-10 mm and 10-15 mm) and four different size ranges (6-8 mm, 8-10 mm, 10-15 mm and 15-20 mm) for sinter and lump ore. Imaging of the grains is done using an iPhone 7 Plus camera and sufficient number of grains of each of the three types are considered for each size range. The imaging methodology is similar to that described in ref. \cite{tripathi} and is briefly described in fig.~\ref{2d_imaging}). 

A total of 1080 particles of a particular size range and type are divided into three sets of 360 particles each. 9 such particles are randomly picked and placed on a white sheet in a non-overlapping fashion. The image of these particles is captured using an iPhone 7 Plus camera using its telephoto lens. To get a high quality picture which is well illuminated from all sides, four LED light bulbs attached on each side of a cardboard box are used. 
The captured image is cropped and converted to RGB format which is then converted to greyscale. Using a threshold value of the pixel intensity, this greyscale image is converted to a black and white image. The black and white colours of resulting image are inverted to finally get a white silhouette of particles on a black background (see last image in fig.~\ref{2d_imaging}). The properties of each silhouette region are obtained using a custom code written in MATLAB utilising the image processing toolbox of the software. For example, the total number of white pixels in each silhouette provide the projected area of the particles in terms of pixel squares. In order to convert this area to mm$^2$, a picture of a 100 mm by 100 mm square area is captured as a reference image in identical imaging conditions. Using the total number of pixels in this reference image, we obtain the appropriate conversion scale relating the length of one pixel to actual physical length. For the results presented in this, the length of a pixel is found to be around a fraction of a millimetre.

Using this pixel-to-mm scale, we calculate various boundary parameters of the silhouette such as the projected area, bounding box length scales, minor and major axis of equivalent ellipse, perimeter, etc. We have defined two additional boundary parameters to characterise the deviation of the projected area from that of a circle and calculate their values for all the different type of grains considered in this study. A brief description of the various silhouette properties calculated using the image analysis are as follows.

\begin{enumerate}
\item Projected Area ($A$): It is measured by calculating the sum of the areas of all pixels in the silhouette region. \item Equivalent Area Diameter ($D_p$): It is defined as the diameter of the circle having an area equal to $A$, i.e., $D_p = \sqrt{\frac{4A}{\pi}}$
\item Mean Radius ($R_{av}$): Defined as $\langle R_i \rangle$, it is average length of the line segments connecting the centroid to different points located on the boundary of the silhouette region.
\item Perimeter ($P$): It is the sum of the distances between each adjoining pair of pixels along the boundary of the region.
\item Major ($2a$) and Minor ($2b$) Axis Lengths: These are the major axis and minor axis lengths, respectively, of an ellipse having the same second moment of inertia as that of the silhouette region.
\item Average deviation length from circular shape ($\sigma$): This parameter is used to quantify the deviation of the projected particle shape from that of a circle. It is calculated as root mean square distance of the difference of length of line segments (such as $OM=R_i$) connecting the center $O$ to any point $M$ located on boundary and the radius ($ON=D_p/2$) of the projected area equivalent circle i.e. $\sqrt{\dfrac{(R_i-\frac{D_p}{2})^2}{N}}$.

\begin{figure}[h!]
\begin{center}
\includegraphics[width=0.4\textwidth]{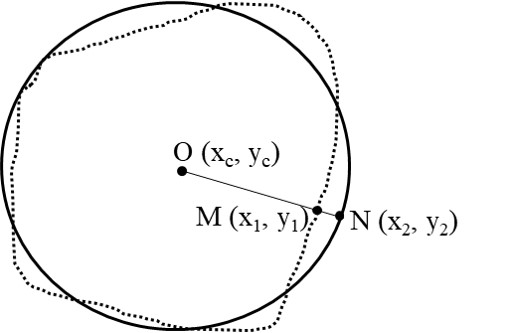}
\caption{Schematic showing the line segments used in the calculation of circular shape deviation length $\sigma$.}
\label{sigma}       
\end{center}
\end{figure}

\item Average Contact Eccentricity ($\langle e\rangle$): The studies by Wensrich et. al. \cite{wensrich, wensrich2} suggest that the ratio of average contact eccentricity $\langle e\rangle$ to the equivalent area diameter $D_p$ appears to be a good estimate of the coefficient of rolling friction $\mu_r$ that can be used in DEM simulations for capturing the rolling difficulty of particles while still representing them as spheres. It is calculated by approximating the silhouette as a polygon by joining points on the boundary and calculating the average distance of the edge bisector normal from the centroid.
Detailed procedure of calculating the same is provided by Agarwal et. al.\cite{agarwal} and Tripathi et.al \cite{tripathi}. 

\item Maximum ($F_{max}$), Minimum ($F_{min}$) and Mean ($F_{mean}$) Feret Diameters: For a given silhouette, a rectangular bounding box encompassing the entire region can be obtained. The dimensions of this bounding box (of length $L_x$ and $L_y$ along the x and y directions) depend upon the orientation of the particle with respect to the coordinate axes. The largest and smallest lengths of the bounding box calculated across all possible orientations are termed as the maximum Feret diameter $F_{max}$ and minimum Feret diameter $F_{min}$, respectively. The average value of all the Feret dimensions of the bounding boxes over all possible orientations is referred to as the mean feret diameter $F_{mean}$.
\item Length ($F_{mbb90}$) and Breadth ($F_{mbb}$) of the Minimum Bounding Box: The bounding box with minimum area is called the minimum bounding box. The length and breadth of this minimum bounding box are calculated as $F_{mbb90}$ and $F_{mbb}$, respectively.
\item Incircle Diameter ($D_i$): It is the diameter of the largest inscribed circle inside the silhouette.
\item Circumcircle Diameter ($D_c$): It is the diameter of the smallest circle circumscribing the silhouette.
\item Convex Hull Perimeter ($P_{ch}$): It is the perimeter of the convex polygon with the least area required to cover the silhouette. 
\item Convex Hull Area ($A_{ch}$): It is the area of the convex polygon with the least area required to cover the silhouette. 
\end{enumerate}

\section{Correlation between the length, perimeter and area scales}

In this section, we attempt to identify the essential silhouette properties that are sufficient to characterize 
the shape of the projected image of different types of particles. The aim is to be able to eliminate the unnecessary measurements a large number of silhouette properties and focus on only the relevant ones for quick and efficient characterisation of the shape descriptors for these materials.
Out of the total 17 silhouette properties calculated using the image analysis in this work, 13 are related to the length scales associated with the particle. Three of these correspond to mean length scales that can be used for characterising particle size. Similarly, we have four possible candidates for minimum length scales and another four related to the maximum
length scales associated with the particle. Two length scales are related to the deviation of the project shape from that of a circle. The remaining four of the seventeen properties correspond to the actual and convex hull perimeter and area. These different types of length scales along with the perimeter scales and area scales are categorised in size different groups and listed in table no.~\ref{length_and_area_scales}. 
The mean, minimum and maximum length scales provide essential information about the overall particle dimensions. Mean length scales can be used to represent the size of the particle, while the minimum and maximum length scales may be useful in representing the aspect ratio of the particles. The deviation length scales from the circle shape scaled with the mean particle size, on the other hand, provide an estimate about the microscopic level details of the grains surface in an averaged sense. The convex hull perimeter and area scaled with their respective true values essentially measure the convexity of the projected shape.
In the following sections, we investigate the correlation of these scale with other scales of the same category and identify the independent scales to reduce the number of possible shape descriptors as much as possible.

\begin{table}[htbp]
\centering
\begin{tabular}{ll}
\hline
Quantity Category & Quantity\\
\hline
\multirow{3}{*}{Mean length scales} & $D_p$\\
& $F_{mean}$\\
& $2R_{av}$\\
\hline
\multirow{4}{*}{Minimum length scales} & $2b$\\
& $D_i$\\
& $F_{min}$\\
& $F_{mbb}$\\
\hline
\multirow{4}{*}{Maximum length scales} & $2a$\\
& $D_c$\\
& $F_{max}$\\
& $F_{mbb90}$\\
\hline
\multirow{2}{*}{Deviation from circular shape length scales} & $\sigma$\\
& $e$\\
\hline
\multirow{2}{*}{Perimeter Scales} & $P$\\
& $P_{ch}$\\
\hline
\multirow{2}{*}{Area scales} & $A$\\
& $A_{ch}$\\
\hline
\end{tabular}
\caption{Length scales, perimeter scales and area scales considered in this study.}
\label{length_and_area_scales}
\end{table}

\subsection{Mean Length Scales}

\begin{figure}[htbp]
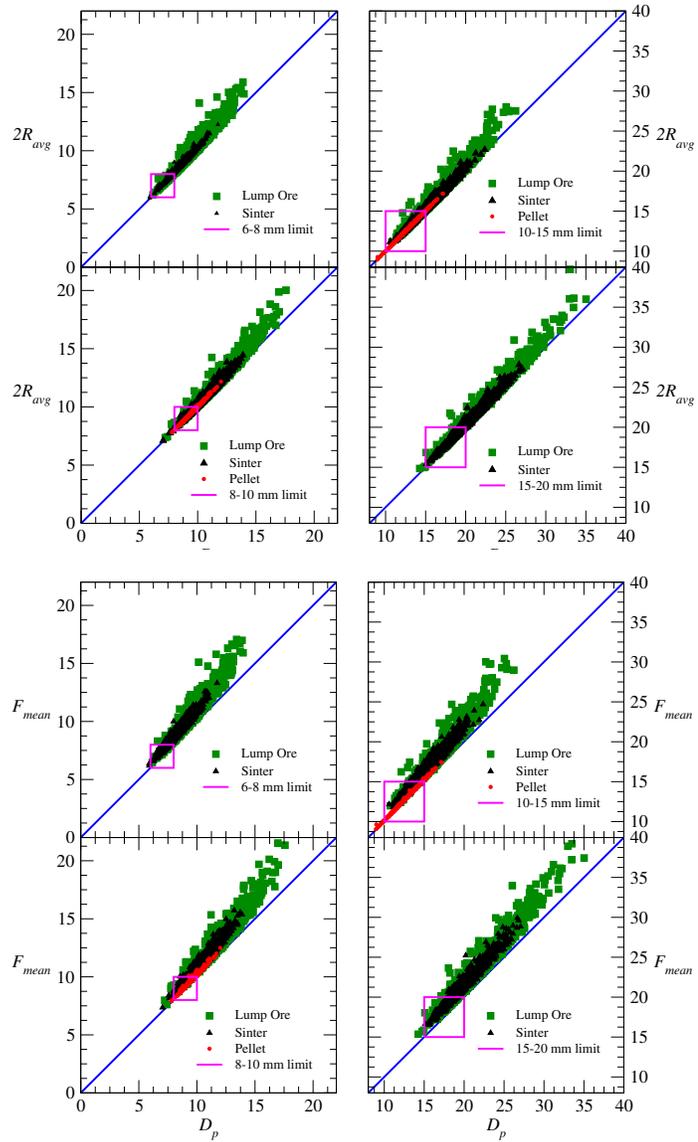

\begin{center}
\includegraphics[width=0.75\textwidth]{Figure_4_1.eps}
\includegraphics[width=0.75\textwidth]{Figure_4_2.eps}
\caption{Relationships between mean length scales. All units in mm. Rectangular box shows the sieve size limit and the solid line corresponds to $y=x$ line.}
\label{mean_lengthscales}       
\end{center}
\end{figure}

Fig.~\ref{mean_lengthscales} shows the relation of $2R_{avg}$ and $F_{mean}$ with $D_p$ for different size ranges of pellet, sinter and lump ore. The solid line corresponds to $y=x$ line and the rectangular box denotes the sieve size range limits. 
For the case of pellets, the relation of the mean lengths closely corresponds to the $y = x$ line, indicating that the three length scales are nearly the same for pellets. This can be attributed to the near-spherical shape of the pellets. 
Though the linear relation holds valid for sinter and lump ore particles as well, slight deviation from $y = x$ line is observable. All the data lie above the $y=x$ line, confirming that both $2R_{avg}$ and $F_{mean}$ are larger than $D_p$. It is also clear that the range of the calculated mean length scales is larger than the sieve size ranges and only a fraction of the data points lie within the sieve size limits. The variation in the mean lengths is largest for lump ore particles and smallest for the pellets.\vspace{5cm}  

\begin{figure}[hbtp]
\begin{center}
\includegraphics[width=0.9\textwidth]{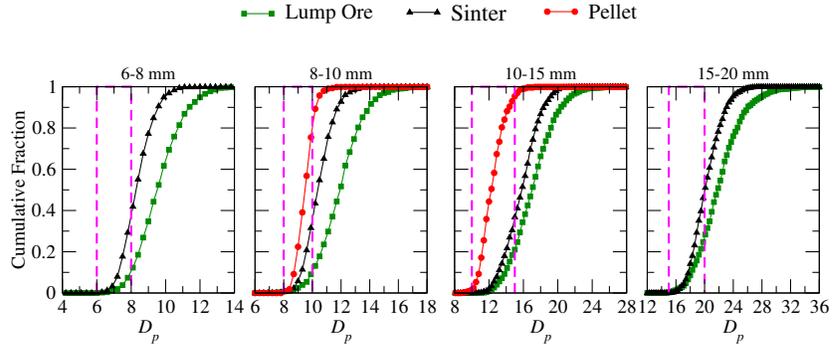}
\caption{Cumulative distribution of $D_p$ (mm). Broken vertical lines show the sieve size limits.}
\label{meanscale_dist}       
\end{center}
\end{figure}

For all the cases reported in this figure, a very strong linear relation of one quantity with the other is observed. The Pearson correlation coefficient for the graphs shown in fig.~\ref{mean_lengthscales} is found to be larger than $0.95$ for all three types of particles over all the size ranges considered. Such a strong correlation coefficient indicates that only one of these three needs to be considered for reporting the mean length scale of the particle. Based on these observations, we select the projected area equivalent circle diameter $D_p$ as the representative mean length scale to quantify particle size for all the three types of materials. 

The variation of $D_p$ for all three types of materials over different size ranges is shown in fig.~\ref{meanscale_dist}. The minimum and maximum sieve size limits are shown using broken vertical lines.
For pellets, middle two panels of fig.~\ref{meanscale_dist}) show that the upper sieve size limit is close to the $90^{th}$ percentile of the data for a size of $8 - 10$ mm, while it is close to the $97^{th}$ percentile when the size range is $10 - 15$ mm. 
The lower sieve size limit is close to the $0^{th}$ percentile for both size ranges, indicating that none of the pellet particles have projected diameters less than the lower sieve size limits. Hence, for the spherical shape pellet particles, the limiting values of the projected diameter closely correspond to the sieve size limits. 
For the sinter particles (shown by triangles in fig.~\ref{meanscale_dist}), whose shape deviates significantly from that of a sphere, the lower sieve size limits seem to coincide with the $0^{th}$ percentile of the $D_p$. The upper sieve size limit, however, corresponds to around $45^{th}$ percentile of the $D_p$ data. For the lump ore, the upper sieve size limit corresponds to the $10^{th}$ percentile of $D_p$ for size ranges of $6 - 8$ and $8 - 10$ mm. However, the upper sieve size limit corresponds to $20^{th}$ percentile of $D_p$ for $10 - 15$ mm and $30^{th}$ percentile for $15 - 20$ mm size lump ore particle. 
These observations may be useful in estimating the upper sieve size limits of the particles of different types using image analysis.

\subsection{Minimum Length Scales}

\begin{figure}[htbp]
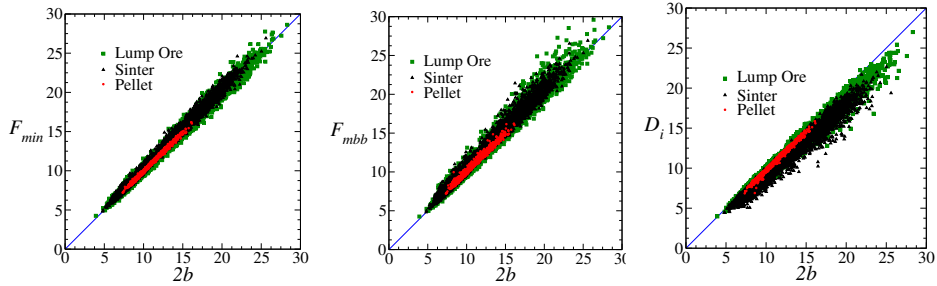

\begin{center}
\includegraphics[width=0.33\textwidth]{Figure_6_1.eps}
\includegraphics[width=0.33\textwidth]{Figure_6_2.eps}
\includegraphics[width=0.32\textwidth]{Figure_6_3.eps}
\caption{Relationships between minimum length scales. All units in mm. Solid line shows the $y=x$ line.}
\label{min_lengthscales}       
\end{center}
\end{figure}

Fig.~\ref{min_lengthscales} shows the relation of $F_{min}$, $F_{mbb}$ and $D_i$ with $2b$ for pellet, sinter and lump ore for all the size ranges together. For pellets, the relationship of these three minimum length scales with $2b$ is linear, closely corresponding to $y = x$ line. The Pearson correlation coefficient for these three length scales with $2b$ and with each other is found to be larger than $0.98$ for pellets. 
Even for sinter and lump ore, $F_{min}$ and $F_{mbb}$ show a strong linear relationship with $2b$ with a value of the Pearson correlation coefficient of $0.92$. Note that the value of $D_i$ is found to be slightly smaller than $2b$ since most of the data points lie below the $y=x$ line.\vspace{5cm}

The Pearson correlation coefficients for $D_i$ with the other $3$ length scales are found to be between $0.8$ and $0.9$. Results for lump ore are also found to be very similar to sinter.  
The results shown in fig.~\ref{min_lengthscales} indicate that $F_{min}$ and $F_{mbb}$ are strongly correlated with  $2b$. While the correlation coefficient of the inscribed circle diameter $D_i$ with $2b$ is slightly smaller compared to that of $F_{min}$ and $F_{mbb}$, its value is still sufficiently large ($>0.8$) to consider the correlation strong. 
Hence, considering only one of these four minimum length scales should be sufficient. Out of these four length scales, the calculation of $F_{min}$, $F_{mbb}$ and $D_i$ requires usage of customised computer programs. The calculation of $2b$, however, can be easily done using the standard image analysis toolbox. Due to this reason, the choice of $2b$ is preferred over the other three choices for minimum length scales. \vspace{5cm} 

\vspace{5cm}
\begin{figure}[h!]
\begin{center}
\includegraphics[width=0.9\textwidth]{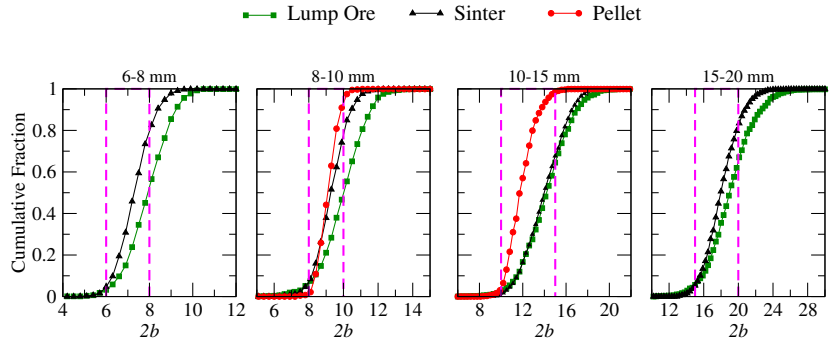}
\caption{Cumulative distribution of $2b$ (mm). Broken vertical lines show the sieve size limits.}
\label{minscale_dist}
\end{center}
\end{figure}

Fig.~\ref{minscale_dist} shows the distribution of minimum length scale $2b$ over different size ranges for pellets, sinter and lump ore. Almost entire data for $2b$ is accommodated in the given sieve sieve limits for pellets, indicating that minimum dimension of the pellet particles also lies with in the sieve size limits.  
However, this is not the case for sinter and lump ore. The upper sieve limits correspond to around $75^{th}$ percentile of the $2b$ data for sinter. For lump ore, it is around $50^{th}$ for $6-8$ mm and $8-10$ mm size ranges and around $60^{th}$ percentile for $10-15$ mm and $15-20$ mm particles.

\subsection{Maximum Length Scales}

\begin{figure}[htbp]
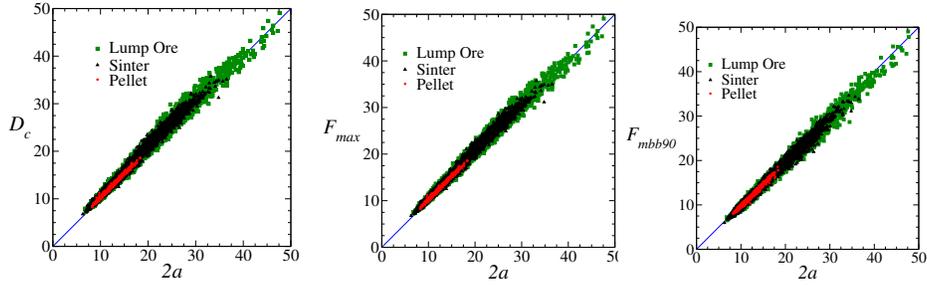

\begin{center}
\includegraphics[width=0.32\textwidth]{Figure_8_1.eps}
\includegraphics[width=0.33\textwidth]{Figure_8_2.eps}
\includegraphics[width=0.33\textwidth]{Figure_8_3.eps}
\caption{Relationships between maximum length scales. All units in mm. Solid line corresponds to $y=x$ line.}
\label{max_lengthscales}       
\end{center}
\end{figure}

Fig.~\ref{max_lengthscales} shows the relation of $D_c$, $F_{max}$ and $F_{mbb90}$ with $2a$ for pellet, sinter and lump ore. For all the cases, these $3$ length scales show a strong linear relationship with $2a$ and the data seems to follow $y = x$ line reasonably well. The Pearson correlation coefficient for all length scale pairings for all the three types of particles is found to be larger than $0.93$, confirming a strong correlation. From this, we can conclude that measuring any one of $F_{max}$, $D_c$, $2a$ and $F_{mbb90}$ is sufficient to characterize the maximum length scale associated with the particle. Due to the availability of standard calculations in image processing toolbox, we select $2a$ as the length scale characterising maximum particle distance.\vspace{5cm} 

\vspace{5cm} 

\begin{figure}[h]
\begin{center}
\includegraphics[width=0.9\textwidth]{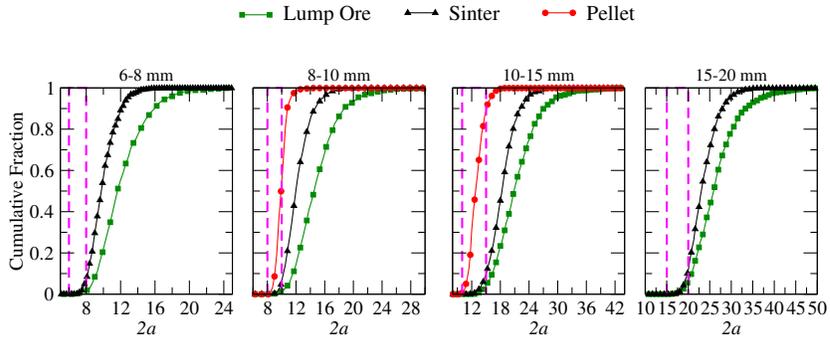}
\caption{Cumulative distribution of $2a$ (mm). Broken vertical lines show the sieve size limits.}
\label{maxscales_dist}       
\end{center}
\end{figure}

The distribution of $2a$ is shown in fig.~\ref{maxscales_dist}. For pellets the upper sieve limit corresponds to about the $90^{th}$ percentile for the size range $10 - 15$ mm (3rd Panel). Since the distribution of $2a$ for size range $8 - 10$ mm (2nd Panel) is quite steep, the upper sieve limit that appears to be about $60^{th}$ percentile though is also not very far from the $90^{th}$ percentile limit. 
The upper sieve size limits correspond to very low percentiles of $2a$ in case of sinter (around $15^{th}$ percentile) and lump ore ($5^{th}$ percentile). This indicates that the largest dimension of the sinter and lump ore particles is much larger than the upper sieve size limit.

\subsection{Observations for different types of perimeter and area measurements} 

\begin{figure}[htbp]
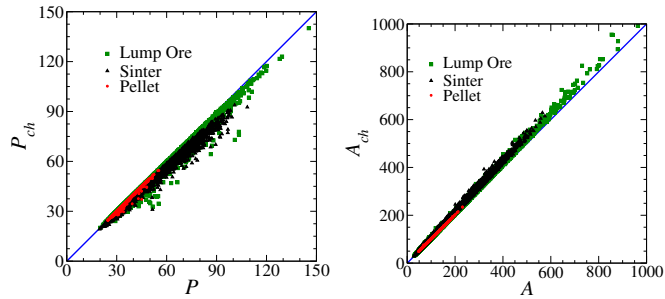

\begin{center}
\includegraphics[width=0.35\textwidth]{Figure_10_1.eps}
\includegraphics[width=0.35\textwidth]{Figure_10_2.eps}
\caption{Variation of the perimeter and area of the particles with that of the convex hull. Units of perimeter and area scales in mm and {mm}$^2$ respectively.}
\label{perimeter_area_scales}       
\end{center}
\end{figure}

Fig.~\ref{perimeter_area_scales} shows the variation of convex hull perimeter and area with the true perimeter and area of the projected images. The slope of the
mean fitted line to the data suggests that the convex hull perimeter differs only
marginally from the true perimeter. A linear variation of the form $P_{ch}$ $=$ $\alpha P$ with $\alpha$ $\leq$ $1$ and $A_{ch}$ $=$ $\beta A$ with $\beta$ $\geq$ $1$ describes the data for all the cases very well. A lot of researchers have used $P_{ch}/P$ as one important shape descriptor and the average values of this parameter for pellet, sinter and lump ore are found to be $0.99$, $0.94$ and $0.97$ respectively. 
Similarly, the convex hull area is also found to be highly correlated with the actual projected area and the proportionality constant for the pellet, sinter and lump ore are found to be $1.01$, $1.07$ and $1.04$ respectively. Some other possible choices of area and perimeter include using these values corresponding to the minimum bounding box and the model elliptical shape of particles using the minor and major axes obtained from imaging. We find all these measures to be highly correlated with the actual perimeter and area and hence do not include them in the analysis.

\subsection{Intercorrelation between different categories of length scales} 

\begin{figure}[htbp]
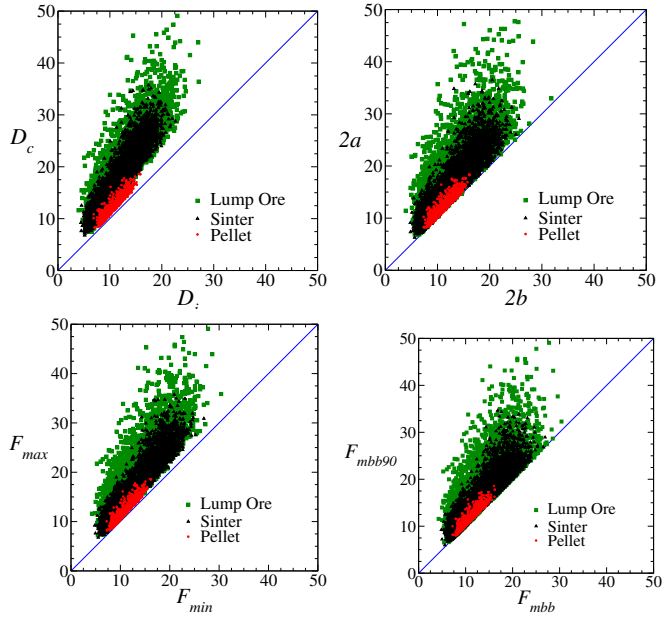

\begin{center}
\includegraphics[width=0.349\textwidth]{Figure_11_1.eps}
\includegraphics[width=0.349\textwidth]{Figure_11_2.eps}
\includegraphics[width=0.349\textwidth]{Figure_11_3.eps}
\includegraphics[width=0.349\textwidth]{Figure_11_4.eps}
\caption{Relationships between maximum and minimum length scales. All units in mm. Solid line corresponds to $y=x$ line.}
\label{maxminscales}       
\end{center}
\end{figure}

The results presented before clearly show that the strong correlation between different particle properties are observable between the same category of particle properties. In this section, we first report data to highlight the important fact  that such high correlation is not observable between properties of two different categories. 
Fig.~\ref{maxminscales} shows the scatter plot of the maximum and minimum length scales plotted against each other. Specifically, we plot $D_c$ vs $D_i$, $2a$ vs $2b$, $F_{max}$ vs $F_{min}$ and $F_{mbb90}$ vs $F_{mbb}$ in Fig.~\ref{maxminscales}. Evidently, the data for these two different class length scales shows significantly larger scatter, especially for the sinter and lump ore particles. The relatively less scatter in case of pellets is attributed to their spherical shape. 
This validates our systematic approach of categorising the particle properties based on their physical relevance. 

\subsection{Final selection of independent particle properties} 

\begin{figure}[htbp]
\begin{center}
\includegraphics[width=0.99\textwidth]{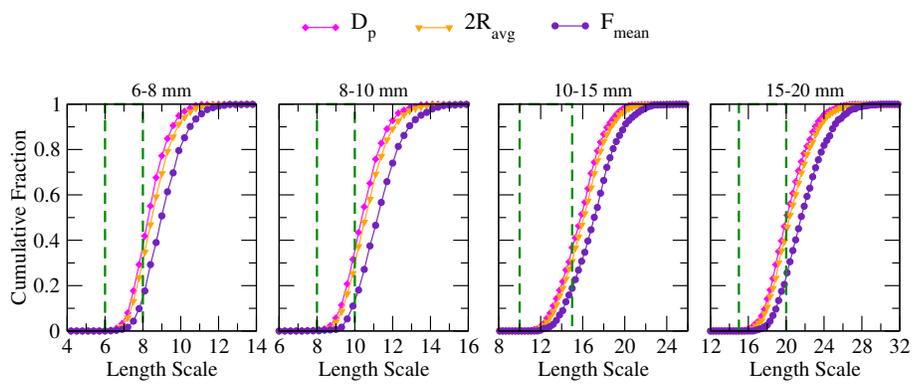}
\caption{Distribution for mean length scales of sinter. All units in mm.}
\label{sinter_meanlengthscales_dist}       
\end{center}
\end{figure}

For gaining better insights about the length scales of the same category, we plot the cumulative distribution curves for the mean length scales of sinter particles in fig.~\ref{sinter_meanlengthscales_dist}. For all the size ranges, the cumulative distribution curves for $D_p$ and $2R_{avg}$ seem to follow each other closely, indicating marginal differences between the two. Further, the distribution curve for $F_{mean}$ is shifted rightwards, confirming our previous observation that $F_{mean}$ is slightly larger than $D_p$ and $2R_{avg}$.\vspace{5cm}

\vspace{5cm} 

\begin{figure}[htbp]
\begin{center}
\includegraphics[width=0.99\textwidth]{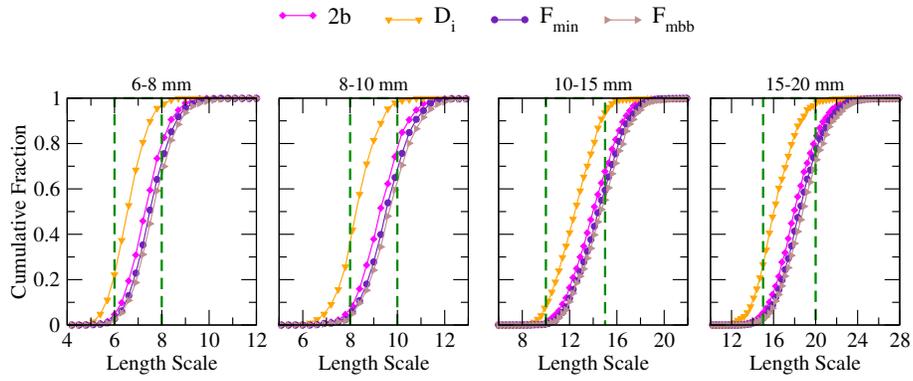}
\caption{Distribution for minimum length scales of sinter. All units in mm.}
\label{sinter_minlengthscales_dist}       
\end{center}
\end{figure}

\begin{figure}[htbp]
\begin{center}
\includegraphics[width=0.99\textwidth]{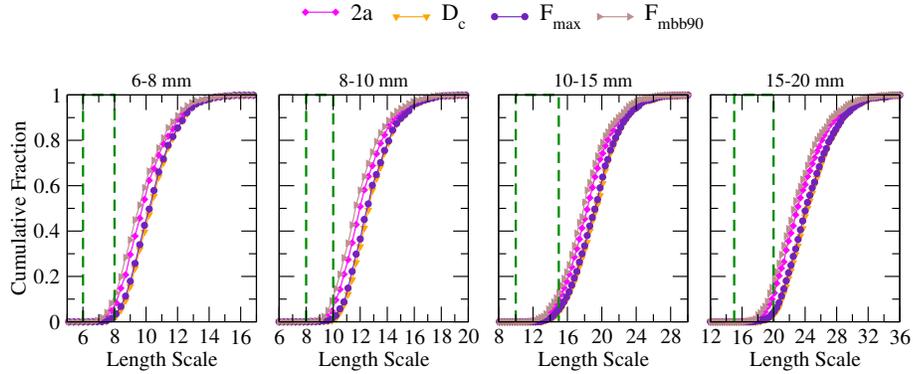}
\caption{Distribution for maximum length scales of sinter. All units in mm.}
\label{sinter_maxlengthscales_dist}       
\end{center}
\end{figure}

Similarly, the distribution of minimum length scales for different size ranges of sinter is plotted in fig.~\ref{sinter_minlengthscales_dist}. The curves for $2b$, $F_{min}$ and $F_{mbb}$ closely follow each other, confirming negligible differences among these three particle properties. The curve for $D_i$, however, is shifted leftwards, confirming our previous observation that $D_i$ is a smaller than the other three minimum length scales.
The cumulative distribution of maximum length scales, plotted in fig.~\ref{sinter_maxlengthscales_dist} confirm that all the four maximum lengths differ little from each other. More specifically, we find that $D_c$ and $F_{max}$ are nearly identical to each other with their values being slightly larger than the nearly identical values of $2a$ and $F_{mbb90}$.\vspace{5cm}

\begin{figure}[htbp]
\begin{center}
\includegraphics[width=0.45\textwidth]{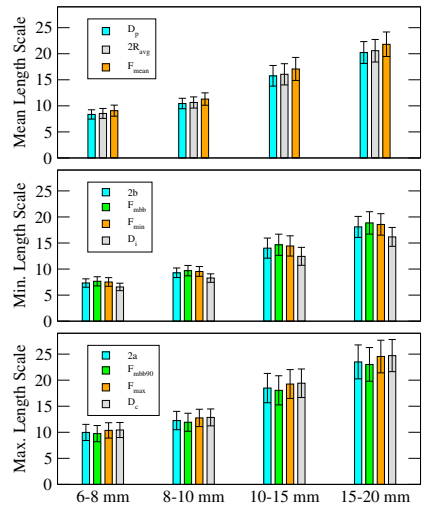}
\caption{Average and Standard deviations of length scales for sinter.}
\label{lengthscales_dist_sinter}       
\end{center}
\end{figure}

\begin{figure}[htbp]
\begin{center}
\includegraphics[width=0.5\textwidth]{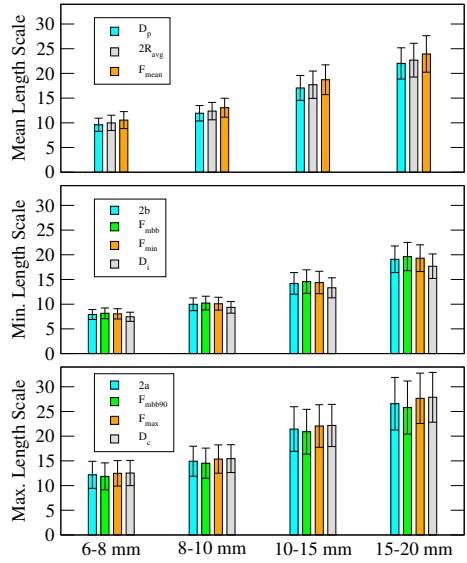}
\caption{Average and standard deviations of length scales for lump ore.}
\label{lengthscales_dist_lump_ore}       
\end{center}
\end{figure}

\begin{figure}[htbp]
\begin{center}
\includegraphics[width=0.4\textwidth]{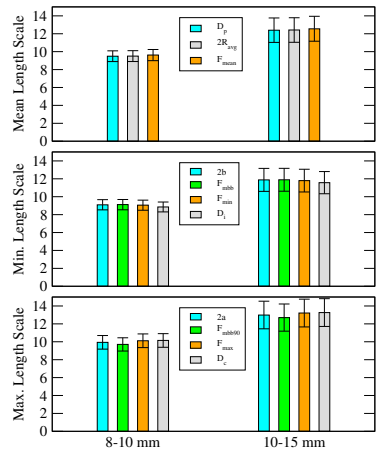}
\caption{Average and standard deviations of length scales for pellets.}
\label{lengthscales_dist_pellet}       
\end{center}
\end{figure}

The near identical values of the length scales corresponding to the mean, minimum and maximum sinter particle dimensions are summarised in fig.~\ref{lengthscales_dist_sinter}. The height of the error bars denotes the standard deviation around the mean values which are represented by the height of the bars in the chart. It is clear from these bar charts that all the length scales of a particular category are very close to each other and the same holds true for the standard deviation. 
Figure~\ref{lengthscales_dist_lump_ore} and \ref{lengthscales_dist_pellet}, show the same data for lump ore and pellets, respectively. Again, it appears that all the properties categorised as the mean length scales are nearly identical to each other with nearly identical standard deviation. The same conclusion holds true for the properties categorised as minimum and maximum length scales associated with the particles as well.\vspace{5cm}  

A few important observations can be made from more a careful comparison of the results presented in this work. These insights can enable the comparison between different length scales used by researchers across different studies. The mean Ferret diameter provides a slightly higher measure of the particle size compared to the equivalent diameter of the projected circle for these non-spherical sinter and lump ore particles. The inscribed circle diameter provides the smallest measure while the width of the minimum area bounding box gives the largest measure of the minimum length scale associated with such particles. Minor axis length of the ellipse with equivalent second moment of inertia provides a value in between these two limits. Similarly, the length of the minimum area bounding box provides the smallest measure of the maximum length scale associated with these particles where as the circumscribing circle diameter, which is almost identical to the maximum Ferret diameter provides the largest measure for the maximum length scale of the particles. The major axis length of the ellipse with equivalent second moment of inertia provides a value in between these two limits. While these conclusions are definitely true for sinter and lump ore, whether they can be extended to general non-spherical shape grains of any kind remains to be seen. \vspace{5cm} 

By using a proper classification methodology of various measured particle properties, coupled with a detailed analysis of the correlation among various possible choices of the same class of length scales, we are eventually able to identify the following independent particle level parameters that need to be considered for calculating the shape descriptors. In addition, we have also included the actual perimeter and projected area of the particles as two independent properties that need to be considered for shape descriptor calculations.

\begin{itemize}
    \item Mean length : Projected Diameter $D_p$ obtained from projected area
    \item Minimum length : Minor axis of equivalent moment of inertia ellipse $2b$
    \item Maximum length : Major axis of equivalent moment of inertia ellipse $2a$
    \item Projected Area $A$
    \item Perimeter $P$
\end{itemize}

\section{Selection of shape descriptors using independent length, perimeter and area scales}

Using the independent and uncorrelated length scales along with the perimeter and area scales identified in the previous section, we can define shape descriptors as the dimensionless combination of these silhouette properties. Since these shape descriptors are essentially ratios of relevant quantities, we select such combinations in a manner so that their values lie between $0$ and $1$.

Table~\ref{sd_length_list} suggests two different choices of the aspect ratio or axial ratio to be quite popular in literature: $b/a$ and $F_{min}/F_{max}$. Some researchers have also used the aspect ratio of the minimum bounding box $F_{mbb}/F_{mbb90}$ for this without explicitly giving any name for it. Our results from the previous section suggest that since $b$, $F_{min}$ and $F_{mbb}$ are strongly correlated to each other and so is the case for $a$, $F_{max}$ and $F_{mbb90}$, these three choices of aspect ratios should be strongly correlated with each other as well. To confirm this, we select the first two, i.e., $b/a$ and $F_{min}/F_{max}$ due to their enormous popularity in literature. 

Very few researchers have also used $F_{mean}/F_{max}$ as a possible candidate for shape descriptor. Table~\ref{sd_length_perimeter_area_list} shows the usage of $2\sqrt A/\sqrt \pi F_{max}$ which is equal to $D_p/F_{max}$ along with the square of this quantity $4A/\pi F^2_{max}=(D_p/F_{max})^2$. These quantities are essentially the ratio of mean to maximum length scales. Recalling our previous results that all the mean as well as maximum length scales are strongly correlated to each other and so is the case for mean lengths, we expect $F_{mean}/F_{max}$ and $D_p/F_{max}$ to be strongly correlated to each other as well as to $D_p/2a$. 
To validate our claim, we consider these three ratios of $F_{mean}/F_{max}$, $D_p/F_{max}$ and $D_p/2a$ as possible candidates for the shape descriptors quantifying the mean to maximum particle length scale ratio.

\begin{table}[htbp]
\centering
\begin{tabular}{cccccc}
\hline

${S}_I = \frac{b}{a}$ & ${S}_{II} = \frac{F_{min}}{F_{max}}$ & ${S}_{III} = \frac{F_{mean}}{F_{max}}$ & ${S}_{IV} = \frac{D_p}{F_{max}}$ \\ \\
${S}_{V} = \frac{D_p}{2a}$ & ${S}_{VI} = \sqrt{\frac{D_i}{D_c}}$ & ${S}_{VII} = \frac{\sigma}{D_p}$ & ${S}_{VIII} = \frac{<e>}{D_p}$ \\ \\
${S}_{IX} = \frac{\pi D_p}{P}$ & ${S}_{X} = \frac{P_{ch}}{P}$ & ${S}_{XI} = \frac{A}{A_{ch}}$ & \\

\hline \\
\end{tabular}
\caption{Shape descriptors selected for study.}
\label{sd_selected}
\end{table}

\begin{table}[htbp]
\centering
\begin{tabular}{cc}
\hline
Shape Descriptor & As a function of selected shape descriptors\\
\hline\\ 

$\frac{\pi F_{mean}}{P}$ & $\frac{\pi D_p}{P} \times (\frac{D_p}{F_{max}})^{-1} \times \frac{F_{mean}}{F_{max}}$\\ \\
\hline \\
$\frac{P}{\pi F_{max}}$ & $(\frac{\pi D_p}{P})^{-1} \times \frac{D_p}{F_{max}}$\\ \\
\hline \\
$\frac{\pi F_{min}}{P}$ & $\frac{\pi D_p}{P} \times (\frac{D_p}{F_{max}})^{-1} \times \frac{F_{min}}{F_{max}}$\\ \\
\hline \\
$\frac{P}{D_p}$ & $\pi (\frac{\pi D_p}{P})^{-1}$\\ \\
\hline \\
$\frac{2 \sqrt{A}}{\sqrt{\pi} F_{mean}}$ & $\frac{D_p}{F_{max}} \times (\frac{F_{mean}}{F_{max}})^{-1}$\\ \\
\hline \\
$\frac{\sqrt{\pi} F_{min}}{2 \sqrt{A}}$ & $(\frac{D_p}{F_{max}})^{-1} \times \frac{F_{min}}{F_{max}}$\\ \\
\hline \\
$\frac{4 A}{\pi {F_{max}}^2}$ & $(\frac{D_p}{F_{max}})^{2}$\\ \\
\hline \\
\end{tabular}
\caption{Shape descriptors used in literature expressed in terms of shape descriptors selected in this study.}
\label{table4}
\end{table}

Table~\ref{sd_length_list} also shows the usage of $\sqrt{(D_i/D_c)}$ as a relevant shape descriptor by some researchers. Based on our analysis of the correlation among the minimum as well as the maximum length scales, this shape descriptor should also be well correlated with the aspect ratio since it is essentially the square root of the length scales ratio that are equivalent to the aspect ratio. To confirm this claim, we select $\sqrt{(D_i/D_c)}$ as another relevant candidate for describing the shape of the particles.
The two other length scales, characterizing the deviation of the projected particle shape from that of a circle, are scaled with $D_p$ to get two additional descriptors $\sigma /D_p$ and $\langle e\rangle/D_p$ that we consider in this study. Since $R_{avg}$ and $D_p$ are correlated, $\sigma$ is correlated with radial shape factor mentioned in Table~\ref{sd_length_list}. As expected, it shows a strong correlation with $b/a$. Another very popular shape descriptor reported in table~\ref{sd_length_perimeter_area_list} is $\pi D_p/P$ and its other variants, such as its inverse $P/\pi D_p$ (also reported in table~\ref{sd_length_perimeter_list}) or the square of these two quantities (see table~
\ref{sd_length_perimeter_area_list}). Given its common occurrence, we select $\pi D_p/P$ as another important shape descriptor in this study. 

The other two most popular shape descriptors reported in table~\ref{sd_length_perimeter_list} and table~
\ref{sd_length_perimeter_area_list} are the ratios of the convex hull perimeter to true perimeter and convex hull area to projected area. Hence, we select $P_{ch}/P$ and $A/A_{ch}$ as the other two important shape descriptors in this study. Thus, a total of 11 shape descriptors are calculated and analysed in this work. These selected shape descriptors are listed in table 5.
Most of the remaining shape descriptors used in literature can be expressed as functions of these shape descriptors (see table~\ref{table4}) and hence are not considered separately. 

\begin{figure}[h]
\begin{center}
\includegraphics[width=0.99\textwidth]{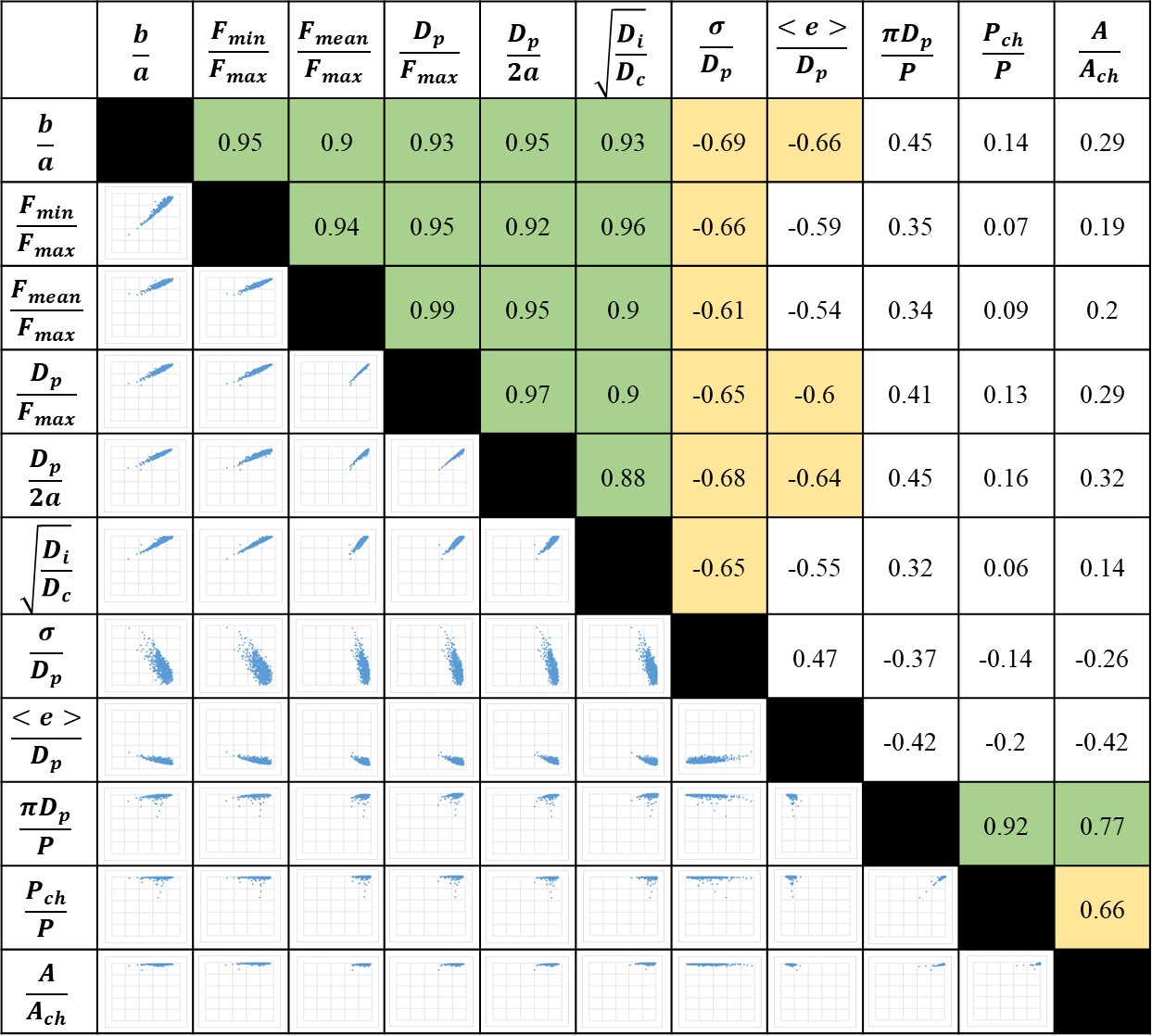}
\caption{Pearson Correlation Coefficients between different shape descriptors for pellets. Graphs in table represent relation between two shape descriptors, corresponding column title being the x-axis and row title being the y-axis.}
\label{pellet_corr}       
\end{center}
\end{figure}

\begin{figure}[h]
\begin{center}
\includegraphics[width=0.99\textwidth]{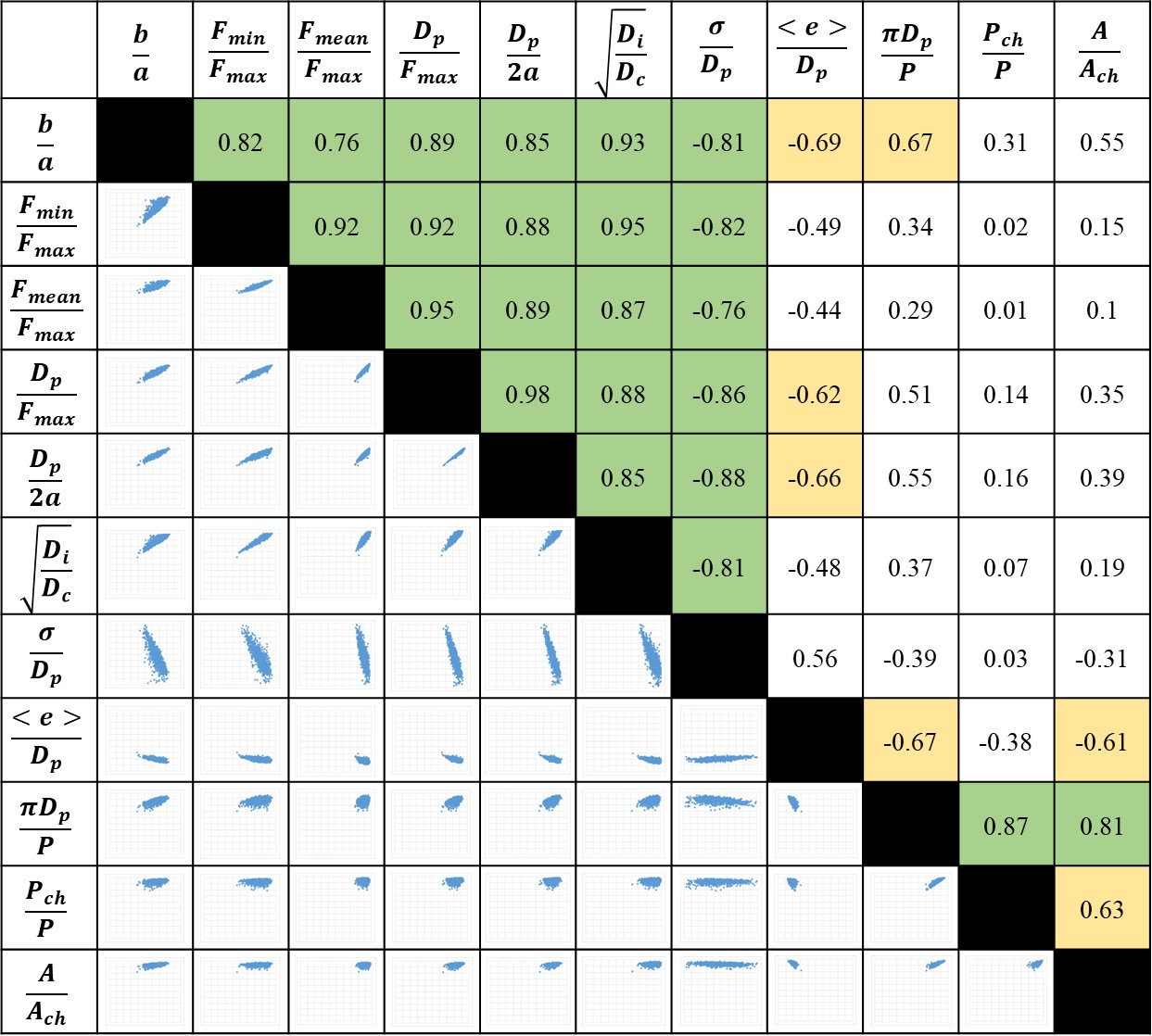}
\caption{Pearson Correlation Coefficients between different shape descriptors for sinter. Graphs in table represent relation between two shape descriptors, corresponding column title being the x-axis and row title being the y-axis.}
\label{sinter_corr}       
\end{center}
\end{figure}

\begin{figure}[h]
\begin{center}
\includegraphics[width=0.99\textwidth]{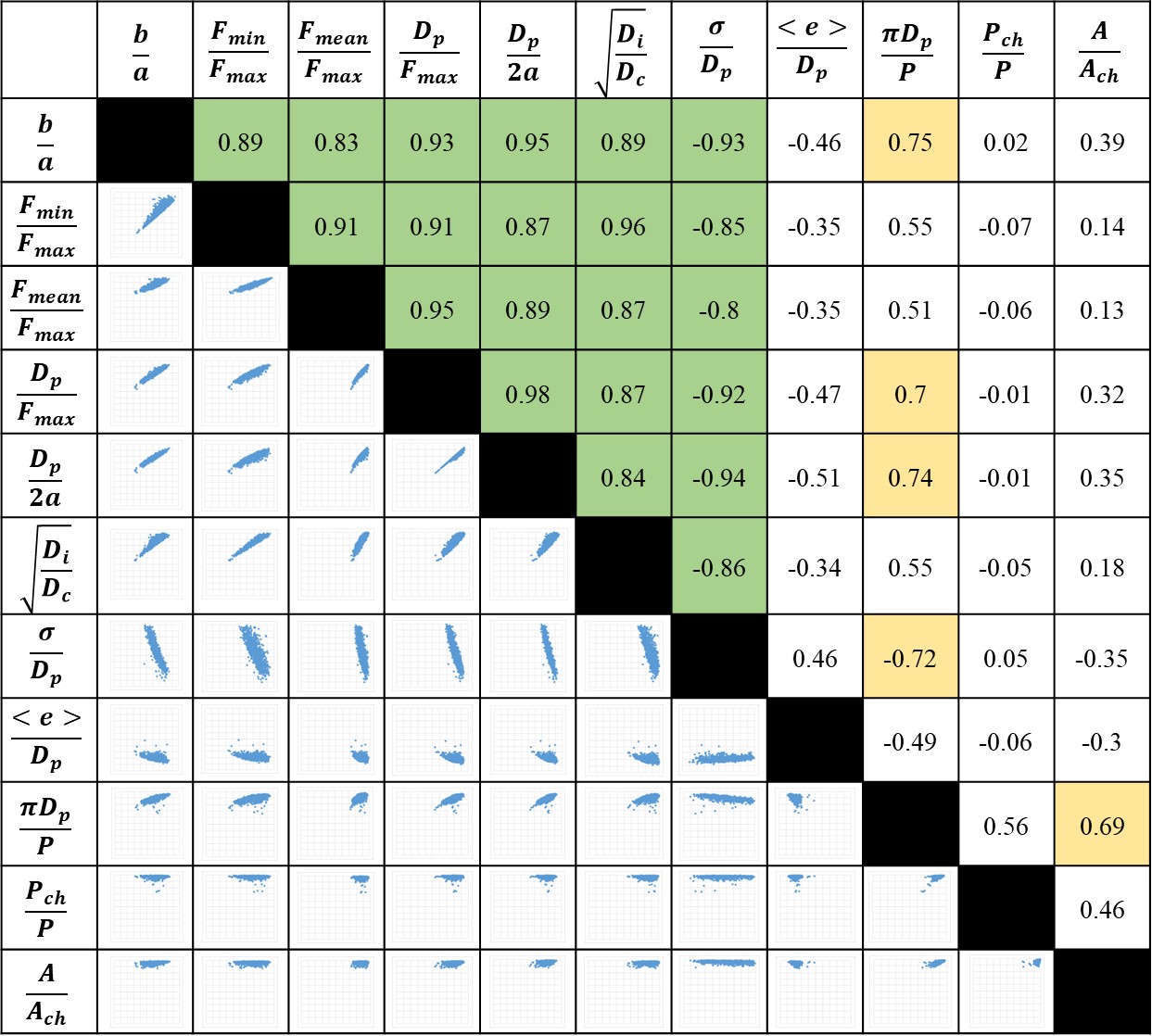}
\caption{Pearson Correlation Coefficients between different shape descriptors for lump ore. Graphs in table represent relation between two shape descriptors, corresponding column title being the x-axis and row title being the y-axis.}
\label{ore_corr}       
\end{center}
\end{figure}

Each of the eleven shape descriptors reported in table~\ref{sd_selected} is correlated with the remaining shape descriptors to obtain a matrix of the Pearson correlation coefficients. Figures \ref{pellet_corr}, \ref{sinter_corr} and \ref{ore_corr} report this correlation matrix for different pairs of shape descriptors for each of the three types of materials, i.e., for pellets, sinter and lump ore, respectively. Due to the symmetric nature of the matrix, we only report the values of the correlation coefficient for the entries above the diagonal of the matrix. The diagonal entries, which report the self correlation coefficients of the shape descriptors (equal to 1) are coloured black due to the trivial information they provide. The entries below the diagonal show the scatter plot of the shape descriptors in a graph. Plots shown in the figure depict relation between two shape descriptors, with the column heading of a particular cell taken as the x-axis and row heading as the y-axis. For all graphs, extent of x-axis and y-axis limits are fixed at $0.5$. The corresponding Pearson correlation coefficient ($R^2$ value) is in the transpose position to that cell. For example, the graph in the cell corresponding to the second column and the third row is $b/a$ vs. $F_{min}/F_{max}$, and the cell corresponding to the third column and second row describes the $R^2$ value for correlation between these two shape descriptors. $|{R^2}|$ $>$ $0.75$ implies strong correlation, and the cells containing those values are colored green. For $|{R^2}|$ between $0.6$ and $0.75$, correlation is considered to be mildly strong and cells with those values colored orange, while $|{R^2}|$ value below $0.6$ implies a weak correlation and are left uncolored.

For all types of particles, shape descriptors $b/a$, $F_{min}/F_{max}$, $F_{mean}/F_{max}$, $D_p/F_{max}$, $D_p/2a$ and $\sqrt{{D_i}/{D_c}}$ show strong correlations with each other. For sinter and lump ore, $\sigma/D_p$ is also strongly correlated with these shape descriptors. However, this correlation for pellets turns out to be only moderately strong, possibly due to the fact that the way $\sigma$ is defined, its value is very small for  nearly spherical shape particles. In this regard, $b/a$ can be considered a unique length scale ratio to quantify particle shape. 
However, $<e>/D_p$, $\pi{D_p}/P$, $P_{ch}/P$ and $A/A_{ch}$ do not seem to be correlated very well to each other as well as with the length scale to length scale ratios mentioned before, despite occasional moderate correlation in some cases. Hence, we select these $4$ alongwith $b/a$ as independent shape descriptors that need to be analysed.

\begin{figure}[h]
\begin{center}
\includegraphics[width=0.99\textwidth]{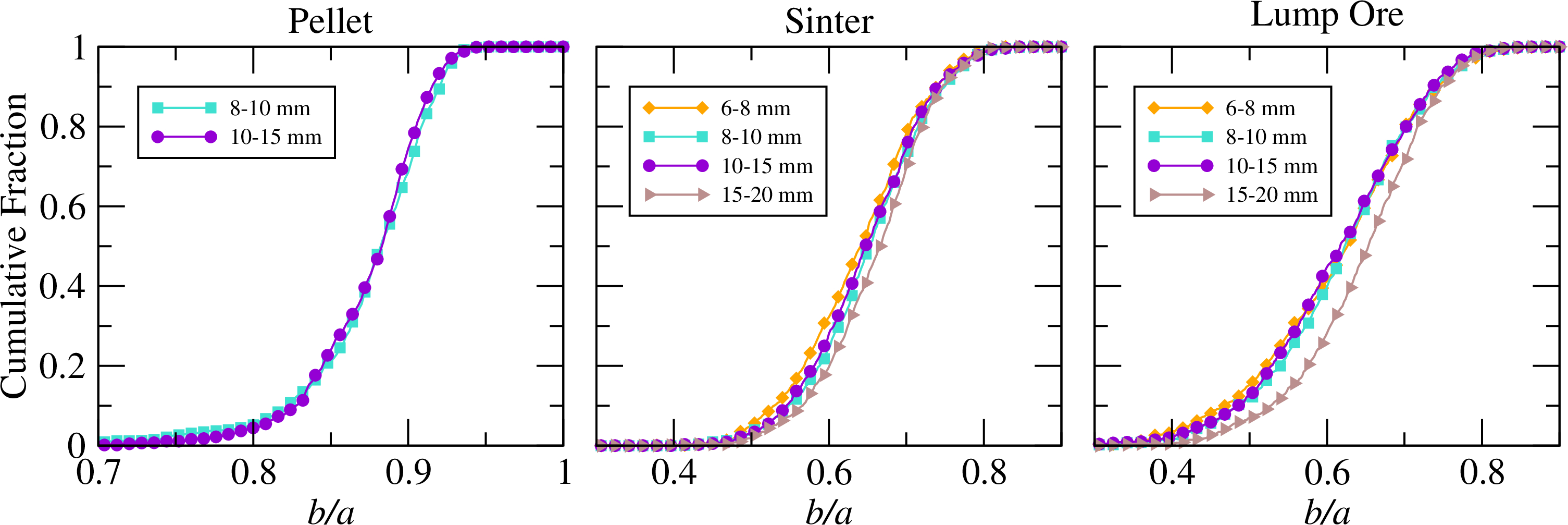}
\caption{Distribution of $b/a$ for different particles.}
\label{b_by_a_dist}       
\end{center}
\end{figure}

\begin{figure}[h]
\begin{center}
\includegraphics[width=0.99\textwidth]{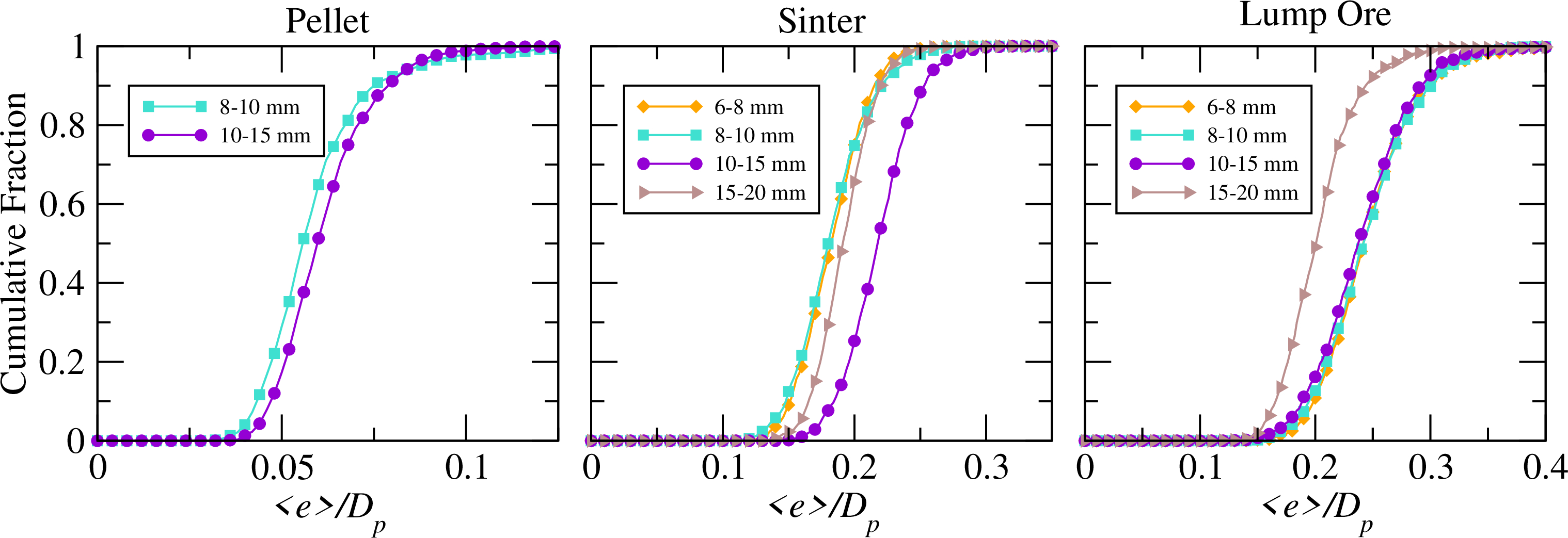}
\caption{Distribution of $<e>/D_p$ for different particles.}
\label{ecc_dist}       
\end{center}
\end{figure}

Fig.~\ref{b_by_a_dist} shows the cumulative distribution of the aspect ratio $b/a$ for different sieve size ranges of each of the pellets, sinter and lump ore particles. For all the three types of materials, the distribution curves for different size ranges are quite close to each other. This indicates that this parameter as the measure of aspect ratio can be used to characterize the shape of all the three types of the particles considered in this study irrespective of their size. Median value (or the $50^{th}$ percentile of the curve) for pellets is around $0.88$. 
For sinter and lump ore, this median value of the aspect ratio is close to $0.65$ and $0.62$ respectively. 
This is further evident from the curves for distribution of $<e>/{D_p}$ (fig.~\ref{ecc_dist}), where median value for pellets is close to $0.06$, implying very low rolling friction for particles due to their near-spherical shape. However, for more irregular shapes like sinter and lump ore, median values for $<e>/{D_p}$ are significantly larger at about $0.2$ and $0.23$ respectively. \cite{tripathi} have shown that this method of obtaining $<e>/{D_p}$ can be used to simulate the effect of non-spherical shape to some extent by using this value as the rolling friction coefficient with spheres of size $D_p$ and can help in reducing the number of parameters that need to be calibrated in discrete element method (DEM) simulations.\newpage

\begin{figure}[h]
\begin{center}
\includegraphics[width=0.99\textwidth]{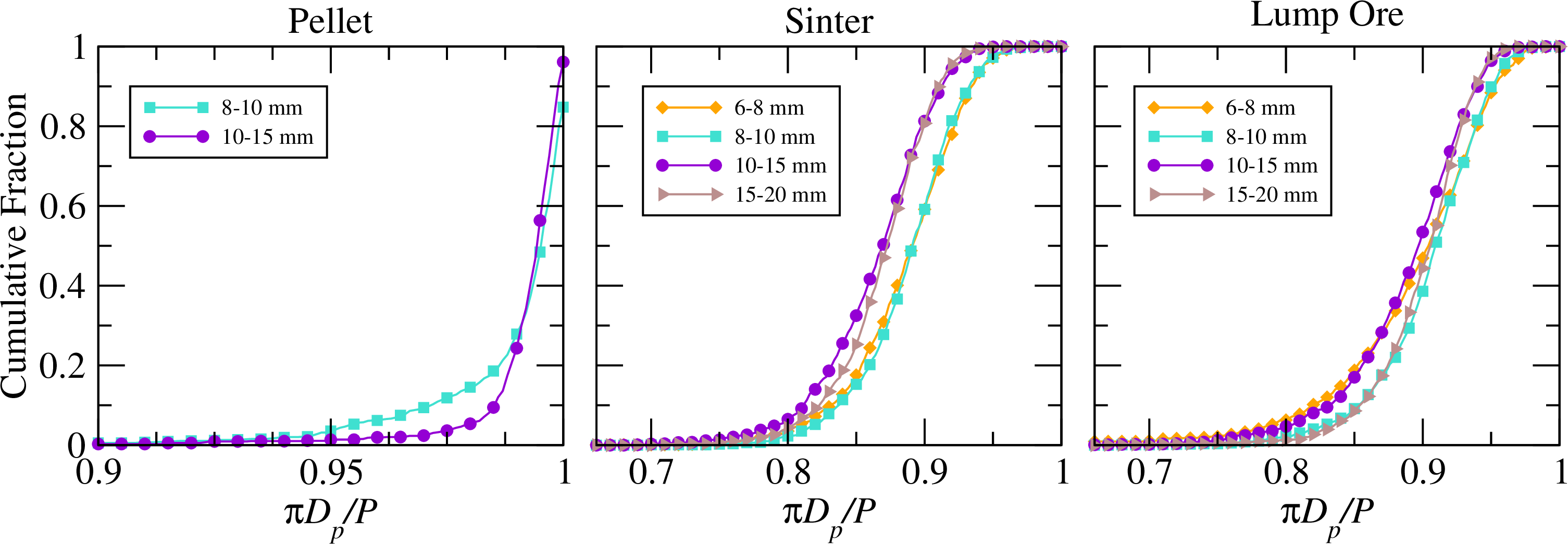}
\caption{Distribution of $\pi D_p/P$ for different particles.}
\label{piDp_by_P_dist}       
\end{center}
\end{figure}

\begin{figure}[h]
\begin{center}
\includegraphics[width=0.99\textwidth]{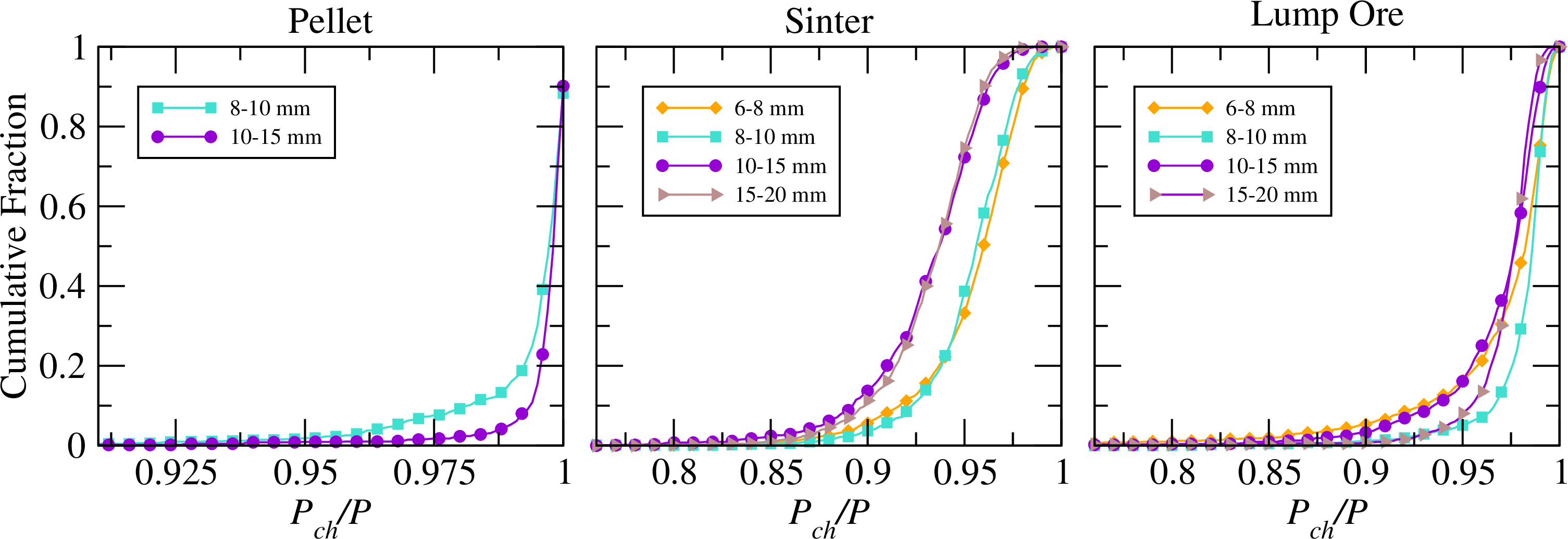}
\caption{Distribution of $P_{ch}/P$ for different particles.}
\label{pch_by_P_dist}       
\end{center}
\end{figure}

\begin{figure}[h]
\begin{center}
\includegraphics[width=0.99\textwidth]{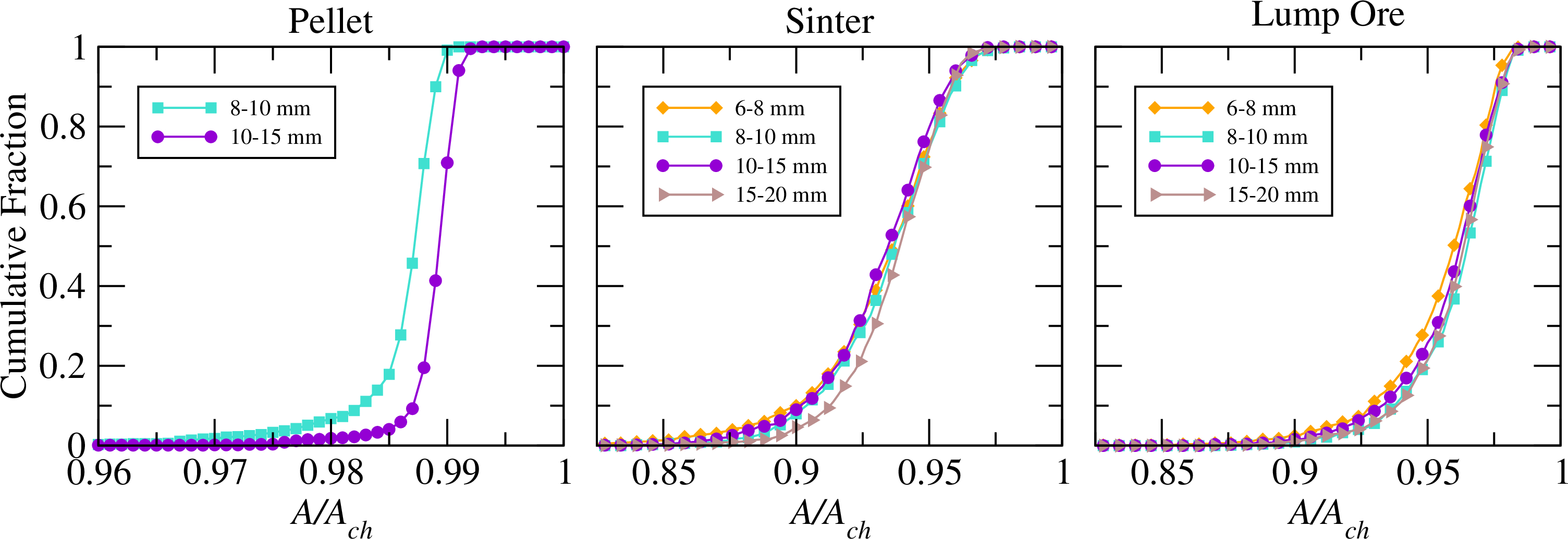}
\caption{Distribution of $A/A_{ch}$ for different particles.}
\label{a_by_ach_dist}       
\end{center}
\end{figure}

Figure~\ref{piDp_by_P_dist} shows the distribution of $\pi D_p/P$ for all the three materials. For pellets, most of the values are greater than $0.97$ due to their near spherical shape. For sinter and lump ore, the median value of $\pi D_p/P$ are found to be around $0.88$ and $0.90$. These close values, despite the fact that sinter edges are quite irregular with several undulations while lump ore has relatively smoother surface, suggest that $\pi D_p/P$ accounts for the overall roundness of the particles by measuring the macroscopic level deviation of the shape from that of a circle and does not account for the edge or surface level smoothness of the particles. Similarly, this difference is also not evident in the distributions of $P_{ch}/P$ and $A/A_{ch}$ either. 
While these values for pellets are quite close to unity (the $10^{th}$ percentile value > $0.98)$, for lump ore, the $10^{th}$ percentile value is greater than $0.93$ for both measures of the particle convexity. For sinter, the $10^{th}$ percentile value for both $P_{ch}/P$ and $A/A_{ch}$ is greater than $0.89$ (see Figs.~\ref{pch_by_P_dist} and \ref{a_by_ach_dist}). 

The detailed values of individual percentile for all shape descriptors and materials of different size ranges is mentioned in table~\ref{sd_selected_values}. As discussed earlier, values of the shape descriptors $P_{ch}/P$ and $A/A_{ch}$ for all materials are fairly close to each other. Further, the $50^{th}$ percentile value is found to be in the range $0.93-0.99$ for all the three types of materials. This is attributed to the fact that the particle contours of the three materials do not seem to exhibit significant concave portions. However, in general, these might turn out to be important for particles that have significantly large undulations with concave portions.

 
The data reported in table ~\ref{sd_selected_values} shows that $b/a$ and $<e>/D_p$ exhibit relatively wider distributions with significant differences between the $10^{th}$, $50^{th}$ and $90^{th}$ percentile values, especially for non-spherical shape sinter and lump ore particles. Hence, these shape descriptors can be used to describe and quantify particle shape in a more reliable fashion. Since the value of $\pi D_p/P$ is quite close to unity for pellets, one may consider representing the same using spheres altogether. For sinter and lump ore, $b/a$ and $<e>/D_p$ need to be considered while defining the shapes along with $\pi{D_p}/P$. The selection of $P_{ch}/P$ and $A/A_{ch}$ values for the blast furnace feed shape can be avoided due to limited surface convexities for the feed particles and little differences of these quantities from unity.


\begin{table}[htbp]
\centering
\begin{tabular}{cccccccc}
\hline
Material & Size Range & ${S}_{(x)
}$ & $\frac{b}{a}$ & $\frac{<e>}{D_p}$ & $\frac{\pi D_p}{P}$ & $\frac{P_{ch}}{P}$ & $\frac{A}{A_{ch}}$\\
\hline
\multirow{6}{*}{Pellet} & \multirow{3}{*}{8-10 mm} & $S_{(10)}$ & $0.82$& $0.04$& $0.97$& $0.98$& $0.98$\\
& & $S_{(50)}$ & $0.88$& $0.06$& $0.995$ & $0.99$ & $0.99$\\
& & $S_{(90)}$ & $0.92$ & $0.07$& $1$ & $1$ & $0.99$\\
\cline{2-8}
& \multirow{3}{*}{10-15 mm} & $S_{(10)}$ & $0.83$& $0.05$& $0.99$& $0.99$& $0.99$\\
& & $S_{(50)}$ & $0.88$& $0.06$& $0.99$& $0.998$ & $0.99$\\
& & $S_{(90)}$ & $0.92$& $0.08$& $0.999$ & $1$ & $0.99$\\
\hline
\multirow{12}{*}{Sinter} & \multirow{3}{*}{6-8 mm} & $S_{(10)}$ & $0.53$& $0.15$& $0.83$ & $0.92$& $0.90$\\
& & $S_{(50)}$ & $0.64$& $0.18$& $0.89$& $0.96$& $0.94$\\
& & $S_{(90)}$ & $0.74$& $0.22$& $0.93$& $0.98$ & $0.96$\\
\cline{2-8}
& \multirow{3}{*}{8-10 mm} & $S_{(10)}$ & $0.55$& $0.15$& $0.84$& $0.92$& $0.90$\\
& & $S_{(50)}$ & $0.65$& $0.18$ & $0.89$& $0.96$ & $0.94$\\
& & $S_{(90)}$ & $0.75$ & $0.22$ & $0.93$ & $0.98$ & $0.96$\\
\cline{2-8}
& \multirow{3}{*}{10-15 mm} & $S_{(10)}$ & $0.55$ & $0.18$ & $0.81$ & $0.89$ & $0.90$\\
& & $S_{(50)}$ & $0.65$ & $0.22$ & $0.87$ & $0.94$ & $0.93$\\
& & $S_{(90)}$ & $0.74$ & $0.25$ & $0.91$ & $0.96$ & $0.96$\\
\cline{2-8}
& \multirow{3}{*}{15-20 mm} & $S_{(10)}$ & $0.56$ & $0.16$ & $0.82$ & $0.90$ & $0.91$\\
& & $S_{(50)}$ & $0.67$ & $0.19$ & $0.87$ & $0.94$ & $0.94$\\
& & $S_{(90)}$ & $0.75$ & $0.22$ & $0.91$ & $0.96$ & $0.96$\\
\hline
\multirow{12}{*}{Lump Ore} & \multirow{3}{*}{6-8 mm} & $S_{(10)}$ & $0.47$ & $0.20$ & $0.82$ & $0.93$ & $0.93$\\
& & $S_{(50)}$ & $0.63$ & $0.24$ & $0.90$ & $0.98$ & $0.96$\\
& & $S_{(90)}$ & $0.74$ & $0.30$ & $0.95$ & $0.99$ & $0.97$\\
\cline{2-8}
& \multirow{3}{*}{8-10 mm} & $S_{(10)}$ & $0.48$ & $0.19$ & $0.85$ & $0.97$ & $0.94$\\
& & $S_{(50)}$ & $0.62$ & $0.24$ & $0.91$ & $0.99$ & $0.96$\\
& & $S_{(90)}$ & $0.74$ & $0.3$ & $0.95$ & $0.99$ & $0.98$\\
\cline{2-8}
& \multirow{3}{*}{10-15 mm} & $S_{(10)}$ & $0.49$ & $0.19$ & $0.83$ & $0.93$ & $0.93$\\
& & $S_{(50)}$ & $0.62$ & $0.24$ & $0.90$ & $0.98$ & $0.96$\\
& & $S_{(90)}$ & $0.74$ & $0.29$ & $0.94$ & $0.99$ & $0.98$\\
\cline{2-8}
& \multirow{3}{*}{15-20 mm} & $S_{(10)}$ & $0.53$ & $0.16$ & $0.85$ & $0.95$ & $0.94$\\
& & $S_{(50)}$ & $0.65$ & $0.20$ & $0.91$ & $0.98$ & $0.96$\\
& & $S_{(90)}$ & $0.75$ & $0.24$ & $0.94$ & $0.99$ & $0.98$\\
\hline
\end{tabular}
\caption{Values of $S_{(10)}$, $S_{(50)}$ and $S_{(90)}$ for selected shape descriptors.}
\label{sd_selected_values}
\end{table}

\section{Conclusion}

Iron ore feed to the blast furnace consists of three different components: pellets, sinter and lump iron ore. 
All three types of particles differ from each other physically as well as chemically. Differences in any of these feed constituents will result in changes in the packing properties of the bed formed in the blast furnace. To understand the effect of the changes in size and shape distribution of these materials on the bed properties, it is important to have a reliable characterization methodology of the feed. In this work, such a methodology is established to characterize the size and shape distribution of the feed material using image analysis. 
We determine various length scales along with perimeter and area scales using MATLAB image analysis toolbox. The length scales are classified into the categories of mean, minimum and maximum length scales along with those representing the deviation of the projected shape from a circle. We find that all length scales within a category are well correlated with each other for sinter, pellet and lump ore. 
This indicates that calculating only one of the length scales in each category is enough to characterize the feed particles. 
We choose equivalent projected area diameter $D_p$, the major axis length $2a$ and minor axis length $2b$ as the three characteristic length scales. 
For pellets, the distribution of these length scales obtained from image processing appears to  lie within the stipulated sieve size limits. However, these length scales are found to be distributed over a much larger range compared to 
the sieve size limits in case of sinter and lump ore particles due to their non-spherical shape.

A comprehensive list of non-dimensional $2$D shape descriptors used in literature is summarised in this work and a couple of new ones have been proposed. From these, $11$ shape descriptors are selected and correlated with each other to find the unique ones that can describe particle shape. As expected, the shape descriptors do not depend upon size range of the particles, confirming that they can indeed be defined irrespective of the particle size. 
Ultimately, three shape descriptors are found to be significantly uncorrelated with the others for all the three types of particles: aspect ratio $b/a$, dimensionless contact eccentricity $<e>/D_p$ and the ratio of the perimeter of equivalent area circle to the true perimeter $\pi {D_p}/P$. The parameter $b/a$ is also referred to as the axial ratio and characterises the ratio of minimum to maximum length scales of the particles. The shape descriptor $\pi {D_p}/P$ is a 2-dimensional equivalent of the sphericity parameter. Recall that sphericity is the ratio of the surface area of equivalent volume sphere to actual surface area. In case of 2D projections, the projected area is used in place of volume and the perimeter is used in place of surface area.
For pellets, values of this shape descriptor is close to unity compared to the other two types of particles. Further, pellets also have the lowest value of $<e>/D_p$. This property has been shown to be equivalent to coefficient of rolling friction that can be utilized in DEM simulations to account for shape effects despite modeling the particles as spheres. Given the values of shape descriptors differ significantly from unity for sinter and iron ore, the non-spherical shape may be important to account for in some situations. 

The systematic approach presented in this work suggests that the image analysis methodology to report shape descriptors can be a good way to characterize and measure the similarities and differences among various feed materials which may arise among different batches due to the variations in their source and operating conditions. Using the three shape descriptors mentioned above, it is possible to reliably describe the shape of iron ore feed used in the blast furnace. The quantitative measure of these shape descriptors can be useful in correlating these shape and size properties to other measurable quantities relevant for furnace operations. 
While some researchers suggest the convex hull perimeter $P_{ch}$ and convex hull area $A_{ch}$ to be important for shape characterization, we find these values to be well correlated with the perimeter $P$ and area $A$ for all the three types of particles. The distribution of $P_{ch}/P$ and $A/A_{ch}$ is found to be rather narrow for sinter and ore and is almost equal to unity in the case of pellets.

\section*{Acknowledgement} {AT gratefully acknowledges the funding support provided by TATA Steel Ltd. for this study.} 
\bibliography{bibfile2}

@article{wensrich,
title = {Rolling friction as a technique for modelling particle shape in {DEM}},
journal = {Powder Technol.},
volume = {217},
pages = {409-417},
year = {2012},
author = {C. M. Wensrich and A. Katterfeld},
doi = {https://doi.org/10.1016/j.powtec.2011.10.057},
}

@article{wensrich2,
title = {Characterisation of the effects of particle shape using a normalised contact eccentricity},
journal = {Granul. Matter},
volume = {16},
pages = {327-337},
year = {2014},
author = {C. M. Wensrich and A. Katterfeld and D. Sugo},
doi = {https://doi.org/10.1007/s10035-013-0465-1},
}

@article{agarwal,
title = {Rolling friction measurement of slightly non-spherical particles using direct experiments and image analysis},
journal = {Granul. Matter},
volume = {23},
number = {3},
pages = {60},
year = {2021},
author = {Arpit Agarwal and Anurag Tripathi and Aman Tripathi and Vimod Kumar and Arijit Chakrabarty and Samik Nag},
doi = {https://doi.org/10.1007/s10035-021-01124-3},
}

@article{hentschelpage,
author = {Hentschel, Mark L. and Page, Neil W.},
title = {Selection of Descriptors for Particle Shape Characterization},
journal = {Part. Part. Syst. Charact.},
volume = {20},
number = {1},
pages = {25-38},
year = {2003},
doi = {https://doi.org/10.1002/ppsc.200390002},
}

@article{bouwmann,
title = {Which shape {factor(s)} best describe granules?},
journal = {Powder Technol.},
volume = {146},
number = {1},
pages = {66-72},
year = {2004},
author = {Anneke M. Bouwman and Jaap C. Bosma and Pieter Vonk and J.(Hans) A. Wesselingh and Henderik W. Frijlink},
doi = {https://doi.org/10.1016/j.powtec.2004.04.044},
}

@article{mikli,
title = {Characterization of Powder Particle Morphology},
journal = {Proceedings of the Estonian Academy of Sciences, Engineering},
volume = {7},
number = {1},
pages = {22-34},
year = {2001},
author = {Valdek Mikli and Helmo Käerdi and Priit Kulu and Michal Besterci},
doi = {https://doi.org/10.3176/eng.2001.1.03},
}

@article{podczeck,
title = {Evaluation of a standardised procedure to assess the shape of pellets using image analysis},
journal = {Int. J. Pharm.},
volume = {192},
number = {2},
pages = {123-138},
year = {1999},
issn = {0378-5173},
author = {F. Podczeck and S. R. Rahman and J. M. Newton},
doi = {https://doi.org/10.1016/S0378-5173(99)00302-6},
}

@inproceedings{olson,
 title={Particle Shape Factors and Their Use in Image Analysis – Part 1 : Theory},
 author={Eric Olson},
 year={2013}
}

@article{berrezueta,
AUTHOR = {Berrezueta, Edgar and {Cuervas-Mons}, José and {Rodríguez-Rey}, Ángel and {Ordóñez-Casado}, Berta},
TITLE = {Representativity of {2D} Shape Parameters for Mineral Particles in Quantitative Petrography},
JOURNAL = {Minerals},
VOLUME = {9},
YEAR = {2019},
NUMBER = {12},
PAGES = {768},
DOI = {https://doi.org/10.3390/min9120768}
}

@article{duris,
title = {Optimal pixel resolution for sand particles size and shape analysis},
journal = {Powder Technol.},
volume = {302},
pages = {177-186},
year = {2016},
author = {Mihal Duriš and Zorana Arsenijević and Darko Jaćimovski and Tatjana {Kaluđerović Radoičić}},
doi = {https://doi.org/10.1016/j.powtec.2016.08.045},
}

@article{lietal,
author = {M. Li and D. Wilkinson and K. Patchigolla},
title = {Comparison of Particle Size Distributions Measured Using Different Techniques},
journal = {Part. Sci. Technol.},
volume = {23},
number = {3},
pages = {265-284},
year  = {2005},
doi = {https://doi.org/10.1016/10.1080/02726350590955912},
}

@article{pintaude,
author = {Pintaude, G. and Coseglio, M.},
title = {Remarks on the application of two-dimensional shape factors under severe wear conditions},
journal = {Friction},
volume = {4},
pages = {65-71},
year  = {2016},
doi = {https://doi.org/10.1007/s40544-016-0105-y},
}

@inbook{endoh,
 author = {Shigehisa Endoh},
 title = {Particle Shape Characterization},
 chapter = {chapter3},
 booktitle = {Powder Technology Handbook},
 publisher = {CRC Press},
 year = {2019},
 doi = {https://doi.org/10.1201/b22268-3},
 }

@article{montero,
author = {Santiago-Montero, Raul and Bribiesca, Ernesto and Santiago, R.},
year = {2009},
month = {1},
pages = {1305-1335},
title = {State of the art of compactness and circularity measures},
volume = {4},
journal = {Int. Math. Forum}
}

@article{bagheri,
title = {On the characterization of size and shape of irregular particles},
journal = {Powder Technol.},
volume = {270},
pages = {141-153},
year = {2015},
author = {G. H. Bagheri and C. Bonadonna and I. Manzella and P. Vonlanthen},
doi = {https://doi.org/10.1016/j.powtec.2014.10.015},
}

@article{grulke,
title = {Size and shape distributions of primary crystallites in titania aggregates},
journal = {Adv. Powder Technol.},
volume = {28},
number = {7},
pages = {1647-1659},
year = {2017},
author = {Eric A. Grulke and Kazuhiro Yamamoto and Kazuhiro Kumagai and Ines Häusler and Werner Österle and Erik Ortel and Vasile-Dan Hodoroaba and Scott C. Brown and Christopher Chan and Jiwen Zheng and Kenji Yamamoto and Kouji Yashiki and Nam Woong Song and Young Heon Kim and Aleksandr B. Stefaniak and D. Schwegler-Berry and Victoria A. Coleman and Asa K. Jämting and Jan Herrmann and Toru Arakawa and Woodrow W. Burchett and Joshua W. Lambert and Arnold J. Stromberg},
doi = {https://doi.org/10.1016/j.apt.2017.03.027},
}

@article{singh,
  title={Powder Characterization by Particle Shape Assessment},
  author={Paramanand Singh and P. Ramakrishnan},
  journal={KONA Powder Part. J.},
  volume={14},
  pages={16-30},
  year={1996},
  doi={https://doi.org/10.14356/kona.1996007}
}

@article{garboczi,
title = {Particle shape and size analysis for metal powders used for additive manufacturing: Technique description and application to two gas-atomized and plasma-atomized {Ti64} powders},
journal = {Addit. Manuf.},
volume = {31},
pages = {100965},
year = {2020},
author = {E. J. Garboczi and N. Hrabe},
doi = {https://doi.org/10.1016/j.addma.2019.100965},
}

@article{michalski,
title = {Application of image analysis for characterization of powders},
journal = {Mater. Sci.-Pol.},
volume = {23},
number = {1},
pages = {79-86},
year = {2005},
author = {J. Michalski and T. Wejrzanowski and R. Peilaszek and K. Konopka and W. Łojkowski and K. {jan Kurzydlowski}},
}

@article{belaroui1,
title = {Wet grinding of gibbsite in a bead-mill},
journal = {Powder Technol.},
volume = {105},
number = {1},
pages = {396-405},
year = {1999},
author = {K. Belaroui and M. N. Pons and H. Vivier and M. Meijer},
doi = {https://doi.org/10.1016/S0032-5910(99)00164-3},
}

@incollection{kulkarni,
title = {Chapter 10 - Microscopy Techniques},
booktitle = {Essential Chemistry for Formulators of Semisolid and Liquid Dosages},
publisher = {Academic Press},
address = {Boston},
pages = {183-192},
year = {2016},
author = {Vitthal S. Kulkarni and Charles Shaw},
doi = {https://doi.org/10.1016/B978-0-12-801024-2.00010-8},
}

@book{wojnar,
author = {Leszek Wojnar},
title = {Image Analysis: Applications in Materials Engineering},
publisher = {CRC Press},
year = {1998},
}

@inproceedings{howe,
author = {P.L. Howard and J. Reintyes and B. J. Roylance and A. Schultz},
title = {New Dimensions in Oil Debris Analysis - The Automated Real-time, On Line Analysis of Debris Particle Shape},
booktitle = {JOAP International Condition Monitoring Conference},
year = {1998},
}

@inproceedings{nowell,
author = {T. Nowell and D. Hodges and B. J. Roylance and T. Barraclough and T. Sperring},
title = {The Development of Technological Support for {RAF} Early Failure Detection Centers},
booktitle = {JOAP International Condition Monitoring Conference},
year = {1998},
}

@article{pons1,
author = {Pons, Marie Noëlle and Vivier, Hervé and Rolland, Thierry},
title = {Pseudo-{3D} Shape Description for Facetted Materials},
journal = {Part. Part. Syst. Charact.},
volume = {15},
number = {2},
pages = {100-107},
year = {1998},
doi = {https://doi.org/10.1002/(SICI)1521-4117(199804)15:2<100::AID-PPSC100>3.0.CO;2-L},
}

@incollection{pons2,
title = {Chapter Fifteen - Particle Shape Characterization by Image Analysis},
booktitle = {Progress in Filtration and Separation},
publisher = {Academic Press},
address = {Oxford},
pages = {609-636},
year = {2015},
author = {Marie Noëlle Pons and John Dodds},
doi = {https://doi.org/10.1016/B978-0-12-384746-1.00015-X},
}

@article{belaroui2,
author = {Belaroui, K. and Pons, M. N.},
year = {2015},
month = {9},
pages = {192},
title = {Method of particle characterisation; morphology by image analysis},
volume = {4},
journal = {J. Fundam. Appl. Sci.},
doi = {https://doi.org/10.4314/jfas.v4i2.9}
}

@article{pons3,
title = {Particle morphology: from visualisation to measurement},
journal = {Powder Technol.},
volume = {103},
number = {1},
pages = {44-57},
year = {1999},
author = {M. N. Pons and H. Vivier and K. Belaroui and B. {Bernard-Michel} and F. Cordier and D. Oulhana and J. A. Dodds},
doi = {https://doi.org/10.1016/S0032-5910(99)00023-6},
}

@article{lecoq,
title = {A grindability test to study the influence of material processing on impact behaviour},
journal = {Powder Technol.},
volume = {105},
number = {1},
pages = {21-29},
year = {1999},
author = {O. Lecoq and P. Guigon and M. N. Pons},
doi = {https://doi.org/10.1016/S0032-5910(99)00114-X},
}

@article{kulu,
title = {Possibilities of evaluation of powder particle granulometry and morphology by image analysis},
journal = {Proc. Estonian Acad. Sci. Eng.},
volume = {4},
pages = {3-17},
year = {1998},
author = {P. Kulu and A. Tumanok and V. Mikli and H. Kaerdi and I. Kohutek and M. Besterci},
}

@book{zeng,
author = {Xian Ming Zheng and Gary Peter Martin and Christopher Marriott},
title = {Particulate Interactions in Dry Powder Formulation for Inhalation},
publisher = {CRC Press},
year = {2001},
doi = {https://doi.org/10.3109/9780203209592},
}

@phdthesis{zhengphd,
author = {Xian Ming Zheng},
title = {The influence of particle engineering on drug delivery by dry powder aerosols},
school = {University of London},
year = {1997},
}

@article{frances,
title = {Particle morphology of ground gibbsite in different grinding environments},
journal = {Int. J. Miner. Process.},
volume = {61},
number = {1},
pages = {41-56},
year = {2001},
author = {C. Frances and N. {Le Bolay} and K. Belaroui and M. N. Pons},
doi = {https://doi.org/10.1016/S0301-7516(00)00025-9},
}

@article{xie,
title = {{3D} size and shape characterization of natural sand particles using {2D} image analysis},
journal = {Eng. Geol.},
volume = {279},
pages = {105915},
year = {2020},
author = {Wei-Qiang Xie and Xiao-Ping Zhang and Xin-Mei Yang and Quan-Sheng Liu and Shao-Hui Tang and Xin-Bin Tu},
doi = {https://doi.org/10.1016/j.enggeo.2020.105915},
}

@article{linzhuli,
title = {Comparison of {2D} and {3D} dynamic image analysis for characterization of natural sands},
journal = {Eng. Geol.},
volume = {290},
pages = {106052},
year = {2021},
issn = {0013-7952},
author = {Linzhu Li and Magued Iskander},
doi = {https://doi.org/10.1016/j.enggeo.2021.106052},
}

@article{maroof,
title = {A new approach to particle shape classification of granular materials},
journal = {Transp. Geotech.},
volume = {22},
pages = {100296},
year = {2020},
author = {Mohammad Ali Maroof and Ahmad Mahboubi and Ali Noorzad and Yaser Safi},
doi = {https://doi.org/10.1016/j.trgeo.2019.100296},
}

@article{mora,
title = {Sphericity, shape factor, and convexity measurement of coarse aggregate for concrete using digital image processing},
journal = {Cem. Concr. Res.},
volume = {30},
number = {3},
pages = {351-358},
year = {2000},
author = {C. F. Mora and A. K. H. Kwan},
doi = {https://doi.org/10.1016/S0008-8846(99)00259-8},
}

@article{yue,
author = {Yue, Zhong Qi and Morin, Isabelle},
title = {Digital image processing for aggregate orientation in asphalt concrete mixtures},
journal = {Can. J. Civ. Eng.},
volume = {23},
number = {2},
pages = {480-489},
year = {1996},
doi = {https://doi.org/10.1139/l96-052},
}

@article{pons4,
author = {Pons, Marie-Noëlle and Vivier, Hervé and Dodds, John},
title = {Particle shape characterization using morphological descriptors},
journal = {Part. Part. Syst. Charact.},
volume = {14},
number = {6},
pages = {272-277},
year = {1997},
doi = {https://doi.org/10.1002/ppsc.19970140603},
}

@article{vivier,
author = {Vivier, Hervé and Marcant, Bruno and Pons, Marie-Noëlle},
title = {Morphological shape characterization: Application to oxalate crystals},
journal = {Part. Part. Syst. Charact.},
volume = {11},
number = {2},
pages = {150-155},
year = {1994},
doi = {https://doi.org/10.1002/ppsc.19940110207},
}

@article{kuo,
author = {C. Kuo and J. D. Frost and J. S. Lai and L. B. Wang},
title = {Three-Dimensional Image Analysis of Aggregate Particles from Orthogonal Projections},
journal = {Transp. Res. Rec.},
volume = {1526},
number = {1},
pages = {98-103},
year = {1996},
doi = {10.1177/0361198196152600112},
}

@article{souza,
title = {Image analysis: Statistical study of particle size distribution and shape characterization},
journal = {Powder Technol.},
volume = {214},
number = {1},
pages = {57-63},
year = {2011},
author = {D. O. C. Souza and F. C. Menegalli},
doi = {https://doi.org/10.1016/j.powtec.2011.07.035},
}

@article{kroner,
title = {Determination of minimum pixel resolution for shape analysis: Proposal of a new data validation method for computerized images},
journal = {Powder Technol.},
volume = {245},
pages = {297-313},
year = {2013},
author = {Stephan Kröner and María Teresa {Doménech Carbó}},
doi = {https://doi.org/10.1016/j.powtec.2013.04.048},
}

@article{becke,
title = {Shape characteristics of suspended solids and implications in different salmonid aquaculture production systems},
journal = {Aquac.},
volume = {516},
pages = {734631},
year = {2020},
author = {Cornelius Becke and Mark Schumann and Juergen Geist and Alexander Brinker},
doi = {https://doi.org/10.1016/j.aquaculture.2019.734631},
}

@article{liu,
title = {Optimising shape analysis to quantify volcanic ash morphology},
journal = {Geo. Res. J.},
volume = {8},
pages = {14-30},
year = {2015},
author = {E. J. Liu and K. V. Cashman and A. C. Rust},
doi = {https://doi.org/10.1016/j.grj.2015.09.001},
}

@inproceedings{he,
author = {Jianqiao He and Houzhen Wei and Jianjun Wang and Huodong Chen},
title = {Evolution of the particle shape of sands under ring shear},
booktitle = {Advances in Engineering Research (AER)},
year = {2017},
}

@proceedings{roostaei,
    author = {Roostaei, Morteza and Soroush, Mohammad and Hosseini, Seyed Abolhassan and Velayati, Arian and Alkouh, Ahmad and Mahmoudi, Mahdi and Ghalambor, Ali and Fattahpour, Vahidoddin},
    title = {Comparison of Various Particle Size Distribution Measurement Methods: Role of Particle Shape Descriptors},
    volume = {Day 2 Thu, February 20, 2020},
    series = {SPE International Conference and Exhibition on Formation Damage Control},
    year = {2020},
    month = {2},
    doi = {https://doi.org/10.2118/199335-MS},
}

@inproceedings{bu,
author = {Bu, Yu Huan and Wei, Hai Ping and Liu, Hua Jie},
title = {The Calculation of Circularity for Floating Beads Based on Image Processing Method of Powder Particles},
year = {2013},
month = {2},
volume = {616},
pages = {2076-2082},
booktitle = {Sustainable Development of Natural Resources},
series = {Advanced Materials Research},
publisher = {Trans Tech. Publications Ltd},
doi = {https://doi.org/10.4028/www.scientific.net/AMR.616-618.2076},
}

@article{mohanty,
title = {Intelligent prediction of engine failure through computational image analysis of wear particle},
journal = {Eng. Fail. Anal.},
volume = {116},
pages = {104731},
year = {2020},
author = {Suvendu Mohanty and Soudip Hazra and Swarup Paul},
doi = {https://doi.org/10.1016/j.engfailanal.2020.104731},
}

@article{zhou,
title = {A Sphere Filling Algorithm for Irregular Aggregate Particle Generation based on Nonlinear Optimization Method},
journal = {KSCE J. Civ. Eng.},
volume = {23},
pages = {120-129},
year = {2019},
author = {Changhong Zhou and Hongzhi Yue and Yuhua Li and Miaomiao Zhang and Jiayin Liu and Shahroz Aijaz},
doi = {https://doi.org/10.1007/s12205-018-0182-8},
}

@article{liu2,
    author = {Liu, E. J. and Cashman, K. V. and Rust, A. C. and Gislason, S. R.},
    title = {The role of bubbles in generating fine ash during hydromagmatic eruptions},
    journal = {Geol.},
    volume = {43},
    number = {3},
    pages = {239-242},
    year = {2015},
    month = {03},
    doi = {https://doi.org/10.1130/G36336.1},
}

@article{liu3,
author = {Liu, E. J. and Cashman, K. V. and Beckett, F. M. and Witham, C. S. and Leadbetter, S. J. and Hort, M. C. and Guðmundsson, S.},
title = {Ash mists and brown snow: {Remobilization} of volcanic ash from recent Icelandic eruptions},
journal = {J. Geophys. Res. Atmos.},
volume = {119},
number = {15},
pages = {9463-9480},
year = {2014},
doi = {https://doi.org/10.1002/2014JD021598},
}

@article{andronico,
title = {Eruption dynamics and tephra dispersal from the 24 {November} 2006 paroxysm at {South-East Crater, Mt. Etna, Italy}},
journal = {J. Volcanol. Geotherm. Res.},
volume = {274},
pages = {78-91},
year = {2014},
author = {Daniele Andronico and Simona Scollo and Maria Deborah {Lo Castro} and Antonio Cristaldi and Luigi Lodato and Jacopo Taddeucci},
doi = {https://doi.org/10.1016/j.jvolgeores.2014.01.009},
}

@article{coltelli,
title = {Characterization of shape and terminal velocity of tephra particles erupted during the 2002 eruption of {Etna volcano, Italy}},
journal = {Bull. Volcanol.},
volume = {70},
pages = {1103–1112},
year = {2008},
author = {M. Coltelli and L. Miraglia and S. Scollo},
doi = {https://doi.org/10.1007/s00445-007-0192-8},
}

@article{manga,
title = {Rounding of pumice clasts during transport: field measurements and laboratory studies},
journal = {Bull. Volcanol.},
volume = {73},
pages = {321-333},
year = {2011},
author = {Michael Manga and Ameeta Patel and Josef Dufek},
doi = {https://doi.org/10.1007/s00445-010-0411-6},
}

@article{nemeth1,
title = {Drivers of explosivity and elevated hazard in basaltic fissure eruptions: The 1913 eruption of {Ambrym Volcano, Vanuatu (SW-Pacific)}},
journal = {J. Volcanol. Geotherm. Res.},
volume = {201},
number = {1},
pages = {194-209},
year = {2011},
author = {Károly Németh and Shane J. Cronin},
doi = {https://doi.org/10.1016/j.jvolgeores.2010.12.007},
}

@article{nemeth2,
author = {Karoly Németh},
title = {Volcanic glass textures, shape characteristics and compositions of phreatomagmatic rock units from the {Western Hungarian} monogenetic volcanic fields and their implications for magma fragmentation},
journal = {Open Geosci.},
number = {3},
volume = {2},
year = {2010},
pages = {399-419},
doi = {https://doi.org/10.1016/10.2478/v10085-010-0015-6},
}

@article{dellino,
title = {Image processing analysis in reconstructing fragmentation and transportation mechanisms of pyroclastic deposits. The case of {Monte Pilato-Rocche Rosse eruptions, Lipari (Aeolian islands, Italy)}},
journal = {J. Volcanol. Geotherm. Res.},
volume = {71},
number = {1},
pages = {13-29},
year = {1996},
author = {P. Dellino and L. {La Volpe}},
doi = {https://doi.org/10.1016/0377-0273(95)00062-3},
}

@article{leibrandt,
title = {Towards fast and routine analyses of volcanic ash morphometry for eruption surveillance applications},
journal = {J. Volcanol. Geotherm. Res.},
volume = {297},
pages = {11-27},
year = {2015},
author = {Sébastien Leibrandt and Jean-Luc {Le Pennec}},
doi = {https://doi.org/10.1016/j.jvolgeores.2015.03.014},
}

@article{jordan,
title = {Processes controlling the shape of ash particles: Results of statistical {IPA}},
journal = {J. Volcanol. Geotherm. Res.},
volume = {288},
pages = {19-27},
year = {2014},
author = {S. C. Jordan and T. Dürig and R. A. F. Cas and B. Zimanowski},
doi = {https://doi.org/10.1016/j.jvolgeores.2014.09.012},
}

@article{durig,
title = {Comparative analyses of glass fragments from brittle fracture experiments and volcanic ash particles},
journal = {Bull. Volcanol.},
volume = {74},
pages = {691-704},
year = {2012},
author = {Tobias Dürig and Daniela Mele and Pierfrancesco Dellino and Bernd Zimanowski},
doi = {https://doi.org/10.1007/s00445-010-0411-6},
}

@article{sinkhonde,
title = {Representativity of morphological measurements and 2-d shape descriptors on mineral admixtures},
journal = {Results Eng.},
volume = {13},
pages = {100368},
year = {2022},
author = {David Sinkhonde and Alladjo Rimbarngaye and Bassirou Kone and Trokon Cooper Herring},
doi = {https://doi.org/10.1016/j.rineng.2022.100368},
}

@article{stienkijumpai,
title = {Development and testing of a novel image analysis algorithm for descriptive evaluation of shape change of a shrinkable soft material},
journal = {Sci. Rep.},
volume = {11},
pages = {18162},
year = {2021},
author = {Pinpinat Stienkijumpai and Maturada Jinorose and Sakamon Devahastin},
doi = {https://doi.org/10.1038/s41598-021-97141-6},
}

@article{tripathi,
title = {Quantitative {DEM} simulation of pellet and sinter particles using rolling friction estimated from image analysis},
journal = {Powder Technol.},
volume = {380},
pages = {288-302},
year = {2021},
doi = {https://doi.org/10.1016/j.powtec.2020.11.024},
author = {Aman Tripathi and Vimod Kumar and Arpit Agarwal and Anurag Tripathi and Saprativ Basu and Arijit Chakrabarty and Samik Nag},
}

@article{kaya,
  title={Particle Shape Modification in Comminution},
  author={Erol Kaya and Richard Hogg and Senthil Kumar},
  journal={KONA Powder Part. J.},
  volume={20},
  pages={188-195},
  year={2002},
  doi={https://doi.org/10.14356/kona.2002021}
}

@misc{imagej,
title = {Image{J}: Image Processing and Analysis in Java},
howpublished = {\url{https://imagej.net/ij/index.html}},
month = {August},
year = {2023},
note = {(Accessed on August 18th, 2023)},
}

@misc{matlab,
author = {{MATLAB}},
title = {Image Processing Toolbox},
howpublished = {\url{https://in.mathworks.com/products/image.html}},
month = {August},
year = {2023},
note = {(Accessed on August 18th, 2023)},
}

@misc{improplus,
author = {{Media Cybernetics}},
title = {Image-Pro},
howpublished = {\url{https://mediacy.com/image-pro/}},
month = {August},
year = {2023},
note = {(Accessed on August 18th, 2023)},
}

@misc{sigmascan,
author = {{Systat Software, Inc.}},
title = {Sigma Scan Pro},
howpublished = {\url{http://www.systat.de/PDFs/sigmascan_brochure.pdf}},
month = {August},
year = {2023},
note = {(Accessed on August 18th, 2023)},
}

@article{han,
title = {A random algorithm for {3D} modeling of solid particles considering elongation, flatness, sphericity, and convexity},
journal = {Comp. Part. Mech.},
volume = {10},
number = {1},
pages = {19–44},
year = {2023},
doi = {https://doi.org/10.1007/s40571-022-00475-9},
author = {Joanne M. R. Fernlund},
}

\end{document}